\documentclass[aps,reprint,amsmath,amssymb,showpacs,showkeys,superscriptaddress]{revtex4-1}
\usepackage{graphicx}
\usepackage{dcolumn}
\usepackage{bm}
\usepackage{hyperref}
\usepackage{natbib}  
\usepackage[justification=justified]{caption}
\usepackage[caption=false]{subfig}
\usepackage{amsopn}
\usepackage{float}
\usepackage{tikz}
\usetikzlibrary{shapes.geometric, arrows}
\tikzstyle{sqr2} = [rectangle, minimum width=3.5cm, minimum height=0.9cm, text centered, draw=black, fill=yellow!30]

\begin{document}

\title[Bayesian semiparametric power spectral density
estimation]{Bayesian semiparametric power spectral density estimation
  with applications in gravitational wave data analysis}


\author{Matthew C. Edwards}
\email{matt.edwards@auckland.ac.nz}
\affiliation{Department of Statistics, University of Auckland, Auckland 1142, New Zealand}
\affiliation{Physics and Astronomy, Carleton College, Northfield, Minnesota 55057, USA}

\author{Renate Meyer}
\email{renate.meyer@auckland.ac.nz}
\affiliation{Department of Statistics, University of Auckland, Auckland 1142, New Zealand}

\author{Nelson Christensen}
\email{nchriste@carleton.edu}
\affiliation{Physics and Astronomy, Carleton College, Northfield, Minnesota 55057, USA}

\begin{abstract}
  The standard noise model in gravitational wave (GW) data analysis
  assumes detector noise is stationary and Gaussian distributed, with
  a known power spectral density (PSD) that is usually estimated using
  clean off-source data.  Real GW data often depart from these
  assumptions, and misspecified parametric models of the PSD could
  result in misleading inferences.  We propose a Bayesian
  semiparametric approach to improve this.  We use a nonparametric
  Bernstein polynomial prior on the PSD, with weights attained via a
  Dirichlet process distribution, and update this using the Whittle
  likelihood.  Posterior samples are obtained using a blocked
  Metropolis-within-Gibbs sampler.  We simultaneously estimate the
  reconstruction parameters of a rotating core collapse supernova GW
  burst that has been embedded in simulated Advanced LIGO noise.  We
  also discuss an approach to deal with non-stationary data by
  breaking longer data streams into smaller and locally stationary
  components.
\end{abstract}

\pacs{04.30.-w, 02.50.-r, 05.45.Tp, 97.60.Bw}


\maketitle

\section{Introduction}

Astronomy is entering a new and exciting era, with the second
generation of ground-based gravitational wave (GW) interferometers
(Advanced LIGO \cite{aLIGO:2015}, Advanced Virgo \cite{aVirgo:2015},
and KAGRA \cite{KAGRA:2013}) expected to reach design sensitivity in
the next few years.  Throughout history, developments in astronomy
have led to deeper understandings of the universe.  Each time we probe
the universe with new sensors, we discover exciting and unexpected
phenomena, that challenge our current beliefs in astrophysics and
cosmology.  GW astronomy promises to do the same, providing a new
set of ears to listen to (potentially unanticipated) cataclysmic
events in the cosmos.

Apart from the first direct observation of GWs, extracting
astrophysical information encoded in GW signals is one of the primary
goals in GW data analysis.  Since observations are subject to noise,
accurate astrophysical predictions rely on an honest characterization
of these noise sources.  At its design sensitivity, Advanced LIGO will
be sensitive to GWs in the frequency band from $10~\mathrm{Hz}$ to
$8~\mathrm{kHz}$.  The main noise sources for ground-based
interferometers include seismic noise, thermal noise, and photon shot
(quantum) noise \cite{aLIGO:2015}.  Seismic noise limits the low
frequency sensitivity of the detectors.  Thermal noise is the
predominate noise source in the most sensitive frequency band of
Advanced LIGO (around $100~\mathrm{Hz}$), and arises from the test
mass mirror suspensions and the Brownian motion of the mirror
coatings.  Photon shot noise is due to quantum uncertainties in the
detected photon arrival rate, and dominates the high frequency
sensitivity of the detectors.

Standard assumptions about the noise model in the GW data analysis
community rely on detector noise being stationary and Gaussian
distributed, with a known power spectral density (PSD) that is usually
estimated using off-source data (not on a candidate signal)
\cite{littenberg:2013}.  Real GW data often depart from these
assumptions \cite{christensen:2010}.  It was demonstrated in
\cite{lsc:2013} that fluctuations in the PSD can moderately bias
parameter estimates of compact binary coalescence GW signals embedded
in LIGO data from the sixth science run (S6).

High amplitude non-Gaussian transients (or ``glitches'') in real
detector data invalidate the Gaussian noise assumption, and
misspecifications of the parametric noise model could result in
misleading inferences and predictions.  A more sophisticated approach
would be to make no assumptions about the underlying noise
distribution by using nonparametric techniques.  Unlike parametric
statistical models, which have a fixed and finite set of parameters
(e.g., the Gaussian distribution has two parameters: $\mu$ and
$\sigma^2$ representing the mean and variance respectively),
nonparametric models have a potentially infinite set of parameters,
allowing for much greater flexibility.

The theory of spectral density estimation requires a time series to be
a stationary process.  If data is not stationary (which is often the
case for real LIGO data), it is important to adjust for this by
introducing a time-varying PSD.  It was demonstrated in
\cite{bayesline:2014} that the noise PSD in real S6 LIGO data is in
fact time-varying.  Variation in detector sensitivity was also shown
in \cite{abadie:2012}.  Other GW literature that discusses
non-stationary noise include \cite{sintes:1998, finn:2001}.  It would
be an over-simplification to assume the Advanced LIGO PSD is constant
over time, and to use off-source data in characterizing this.
On-source estimation of the PSD would therefore be preferable to
mitigate the time-varying nature of the PSD.

There have been attempts reported in the literature to improve the
modelling of noise present in GW data, primarily concentrating on
noise with embedded signals from well-modelled GW sources, such as
binary inspirals \cite{littenberg:2013, roever:2011a, roever:2011b,
  littenberg:2010, vitale:2014}, and more recently from GW
\textit{bursts} (un-modelled and typically short duration events)
\cite{bayesline:2014, bayeswave:2014}.

Under the Bayesian framework, R\"{o}ver \textit{et al}
\cite{roever:2011a} used a Student-$t$ likelihood as a generalization
to the commonly used Whittle (approximate Gaussian) likelihood
\cite{whittle:1957}.  The benefit of the Student-$t$ set-up is
two-fold: uncertainty in the noise spectrum can be accounted for via
marginalization of nuisance parameters; and outliers can be
accommodated due to the heavy-tail nature of the Student-$t$
probability density.  A drawback of this method is that the choice of
hyperparameters can unduly influence posterior inferences.

Using the maximum likelihood approach, R\"{o}ver \cite{roever:2011b}
later demonstrated that the Student-$t$ likelihood could be used as a
generalization to the matched-filtering detection method commonly used
in the analysis of GW signals from well-modelled sources.  This
approach would not be appropriate for GW bursts, since
matched-filtering requires accurate signal models with well-defined
parameter spaces.

Littenberg and Cornish \cite{littenberg:2010} used Bayesian model
selection to determine the best noise likelihood function in
non-Gaussian noise. They considered Gaussian noise with a time varying
mean, noise from a weighted sum of two Gaussian distributions
(non-Gaussian tails), and a combination of Gaussian noise and glitches
(modelled as a linear combination of wavelets).

Littenberg \textit{et al} \cite{littenberg:2013} demonstrated how one
can incorporate additional scale parameters in the Gaussian
likelihood, and marginalize over the uncertainty in the PSD to reduce
systematic biases in parameter estimates of compact binary mergers in
S5 LIGO data.  This method requires an initial estimate of the PSD.
On a related note, Vitale \textit{et al} \cite{vitale:2014}
highlighted a Bayesian method, similar to iteratively reweighted least
squares, that analytically marginalizes out background noise and
requires no \textit{a priori} knowledge of the PSD.  They applied this
to simulated data from LISA Pathfinder.

More recently, Littenberg and Cornish \cite{bayesline:2014} introduced
the BayesLine algorithm in conjunction with BayesWave
\cite{bayeswave:2014} to estimate the underlying PSD of GW detector
noise.  BayesLine is used to model the Gaussian noise component.  They
use a cubic spline to model the smooth changing broadband noise and
Lorentzians (Cauchy densities) to model wandering spectral lines (due
to AC supply, violin modes, etc.). BayesWave, on the other hand,
models the non-Gaussian instrument ``glitches'' and burst sources with
a continuous wavelet basis.  Both methods make use of the
trans-dimensional reversible jump Markov chain Monte Carlo (RJMCMC)
algorithm of Green \cite{green:1995}.  BayesLine is very pragmatic and
works remarkably well on real Advanced LIGO data.  However, the
authors did not consider statistically important notions such as the
posterior consistency of the PSD \cite{BNP:2010}.

Our approach to improving the GW noise model relies on developments
over the past decade in the area of Bayesian nonparametrics.  Since
parametric modelling can lead to biased estimates when the underlying
parametric assumptions are invalid, we prefer nonparametric techniques
to estimate the PSD of a stationary noise time series.

A common nonparametric estimate of the spectral density of a stationary time
series is the periodogram, calculated using the (normalized) squared
modulus of Fourier coefficients.  That is, 
\begin{equation}
  \label{eq:periodogram}
  I_n(\lambda) = \frac{1}{2\pi n} \left| \sum_{t = 1}^n X_t \exp(-\mathrm{i}t\lambda)\right|^2, \quad \lambda \in (-\pi, \pi], 
\end{equation}
where $\lambda$ is the frequency, and $\{X_t\}$ is a stationary time
series, where $t = 1, 2, \ldots, n$ represents discretized time. The
periodogram randomly fluctuates about the true spectral density of a
time series, but is not a consistent estimator, motivating methods
such as periodogram smoothing and averaging \cite{welch:1967}.
Averaging of off-source periodograms from Tukey windowed simulated
Advanced LIGO noise has been used in GW literature relating to
reconstructing rotating core collapse GWs \cite{roever:2009}, and
predicting the important astrophysical parameters from these events
\cite{edwards:2014}.

In this paper, we implement the nonparametric Bayesian spectral
density estimation method and Metropolis-within-Gibbs Markov chain
Monte Carlo (MCMC) sampler presented by Choudhuri \textit{et al}
\cite{choudhuri:2004}, which updates a nonparametric Bernstein
polynomial prior \cite{petrone:1999a, petrone:1999b} on the spectral
density using the Whittle likelihood to make posterior inferences.  A
Bernstein polynomial prior is essentially a finite mixture of Beta
probability densities (see Section~II~C and Appendix~A).  It was
proved that this method yields a consistent estimator for the true
spectral density of a (short-term memory) stationary time series
\cite{choudhuri:2004} --- an attractive feature, absent in the
periodogram.  Posterior consistency in this context essentially means
that the posterior probability of an arbitrary neighbourhood around
the true PSD goes to 1 as the length of the time series increases to
infinity.  Thus, as the sample size increases, the posterior
distribution of the PSD will eventually concentrate in a neighbourhood
of the true PSD \cite{BNP:2010}.  This is an important asymptotic
robustness quality of the posterior distribution in that the choice of
prior parameters should not influence the posterior distribution too
much.  Especially in Bayesian nonparametrics, because of the high
dimension of the parameter space, many posterior distributions do not
automatically possess this quality \cite{BNP:2010}.  We refer the
reader to Appendix~C for a visual demonstration of posterior
consistency.

Unlike references \cite{roever:2011a, littenberg:2013, vitale:2014},
we do not treat noise as a nuisance parameter to be analytically
integrated out.  Although the signal parameters are our primary
concern, we are also interested in quantifying our uncertainty in the
underlying PSD of the noise in terms of posterior probabilities and
credible intervals.  Knowledge of this uncertainty will allow us to
make honest astrophysical statements.

In this study, we assume that data is the sum of a GW signal embedded
in noise (from all noise sources), such that
\begin{equation}
  \label{eq:setup}
  \mathbf{y} = \mathbf{s}(\boldsymbol\beta) + \boldsymbol\epsilon(\boldsymbol\theta),
\end{equation}
where $\mathbf{y}$ is the (coincident) time-domain GW data vector,
$\mathbf{s}$ is a GW signal parameterized by $\boldsymbol\beta$, and
$\boldsymbol\epsilon$ is noise parameterized by $\boldsymbol\theta$.
Notation with a tilde on top, such as $\tilde{\mathbf{y}}$, refers to
the frequency-domain equivalent of the same quantity, obtained by the
discrete Fourier transform (DFT).  Note that we are treating noise in
this set-up as the conglomeration of detector noise (such as thermal
noise and photon shot noise), background noise (such as seismic
noise), and residual errors due to parametric statistical modelling of
GW signals.  An important caveat is to ensure the magnitude of the
errors in the statistical model of the signal are minimized, so as to
not artificially dominate the noise.  Estimation of spectral lines (as
done by the BayesLine algorithm \cite{bayesline:2014}) is left out of
the scope of this paper.

The GW signal could essentially come from any source, but in this
paper we will restrict our concentration to those from rotating core
collapse supernovae to simplify the problem and demonstrate the power
of the method.  Using the recent waveform catalogue of Abdikamalov
\textit{et al} \cite{abdikamalov:2014}, we conduct principal component
analysis (PCA), and fit a principal component regression (PCR) model
of the most important principal components (PCs) on an arbitrary
rotating core collapse GW signal \cite{roever:2009, edwards:2014,
  heng:2009}.  The (parametric) signal component is easily embedded as
an additional Gibbs step in the Metropolis-within-Gibbs MCMC sampler
of Choudhuri \textit{et al} \cite{choudhuri:2004}.  That is, we
utilize a \textit{blocked} Gibbs approach to sequentially sample the
signal parameters $\boldsymbol\beta$ given the noise parameters
$\boldsymbol\theta$, and vice versa.  As the model now contains a
parametric signal component as well as a nonparametric noise
component, it is ``semiparametric''.

To accommodate for non-stationary noise, we adapt an idea presented by
Rosen \textit{et al} \cite{rosen:2012}, and assume a non-stationary
time series can be broken down into smaller locally stationary
segments.  For each segment, we separately estimate the PSD using the
method of Choudhuri \textit{et al} \cite{choudhuri:2004}, and look at
the time-varying spectrum.

We see this work as being a complement to existing methods, with the
following benefits:
\begin{itemize}
\item A Bayesian framework, allowing us to update prior knowledge based
  on observed data, as well as quantify uncertainty in terms of probabilistic statements;
\item Posterior consistency of the PSD, i.e., the posterior
  distribution will concentrate around the true PSD as the sample size
  increases;
\item No parametric assumptions about the underlying noise
  distribution (parametric models are very sensitive to
  misspecifications), and high amplitude non-Gaussian transients in
  the noise can be handled;
\item Non-stationarities can be taken into account by splitting the data into
  smaller locally stationary segments;
\item Estimation of noise and signal parameters are done
  simultaneously using Gibbs sampling;
\item Uncertainty in astrophysically meaningful parameter
  estimates are honest, with less systematic bias present;
\item Non-informative priors can be chosen, and the
  PSD does not need to be known \textit{a priori};
\item Useful for any signal with a parametric statistical model
  (including rotating core collapse supernova GWs).
\end{itemize}

The paper is structured as follows: Section II outlines the methods
and models used to simultaneously estimate signal and noise parameters
in GW data; results for toy models and simulated Advanced LIGO data
are presented in Section III; and in Section IV, we discuss the
consequences of this work, as well as future initiatives.
Supplementary material can be found in the three appendices.

\section{Methods and Models}

\subsection{Parametric, Nonparametric, and Semiparametric Models}

Statistical models can be classified into two groups --- parametric
and nonparametric.  \textit{Parametric} models have a fixed and finite
set of parameters, are relatively easy to analyze, and are powerful
when their underlying assumptions are correctly specified.  However,
if the model is misspecified, inferences will be unreliable.
\textit{Nonparametric} models have far fewer restrictions, but are
less efficient and powerful than their parametric counterparts.  No
assumption about the underlying distribution of the data is made in
nonparametric modelling, and the number of parameters are not fixed
(and potentially infinite dimensional).  Instead, the effective number
of parameters increases with more data, providing the model structure.

For example, parametric regression (including linear models, nonlinear
models, and generalized linear models) uses the following equation:
\begin{equation}
  \label{eq:parametric}
  \mathbf{y} = g(\mathbf{x_1, x_2, \ldots, x_k} | \boldsymbol\beta) + \boldsymbol\epsilon,
\end{equation}
where $\mathbf{y}$ is the response variable, $g(\mathbf{x_1, x_2,
  \ldots, x_k} | \boldsymbol\beta)$ is a function of $k$ explanatory
variables (that aim to explain the variability in $\mathbf{y}$) given
some model parameters $\boldsymbol\beta$, and $\boldsymbol\epsilon$ is
the statistical error, usually assumed to be independent and
identically distributed (iid) Gaussian random variables, with 0 mean and
constant variance $\sigma^2$.  Here, the functional form of $g(.)$ is
known in advance, such as in linear regression, where we have
\begin{equation}
  \label{eq:linear}
  g(\mathbf{x_1, x_2, \ldots, x_k} | \boldsymbol\beta) = \beta_0 +
\beta_1\mathbf{x_1} + \ldots + \beta_k\mathbf{x_k}.
\end{equation}
Nonparametric regression has a similar set-up, but assumes that the
functional form of $g(.)$ is unknown and to be learnt from the data.
Function $g(.)$ could be thought of as an uncountably
infinite-dimensional parameter in a nonparametric setting.

\textit{Semiparametric} models contain both parametric and
nonparametric components.  The parametric regression model presented
in Equations~(\ref{eq:parametric})~and~(\ref{eq:linear}) is
essentially the same parametric model used in this paper for GW
signal reconstruction, where $(\mathbf{x_1, x_2, \ldots, x_k})$ are
principal component (PC) basis functions.  However, we model the noise
$\boldsymbol\epsilon$ nonparametrically, rather than assuming iid
Gaussian noise.  Since we have parametric and nonparametric
components, our model is semiparametric in nature.

\subsection{Bayesian Nonparametrics}

Bayesian nonparametrics contains the set of models on the interface
between the Bayesian framework and nonparametric statistics, and is
characterized by large parameter spaces and probability measures over
these spaces \cite{BNP:2010}.  The Bayesian statistical framework is
useful for incorporating prior knowledge, and is particularly powerful
when these priors accurately represent our beliefs.  As mentioned in
the previous section, nonparametric methods are useful for
constructing flexible and robust alternatives to parametric models.  A
benefit of Bayesian nonparametric models is that they automatically
infer model complexity from the data, without explicitly conducting
model comparison.

Bayesian nonparametrics is a relatively nascent field in statistics,
and faces many challenges.  The most obvious one is the mathematical
difficulty in specifying well-defined probability distributions on
infinite-dimensional function spaces.  Constructing a prior on these
spaces can be arduous, and in the case of non-informative priors, one
should ensure large topological support so as not to put too much mass
on a small region.  Further, creating computationally convenient
algorithms to sample from complicated posterior distributions presents
its own set of challenges.  It is also important to ensure that a
Bayesian nonparametric model is statistically consistent (the truth is
uncovered asymptotically), as some procedures do not automatically
possess this quality \cite{BNP:2010}.

Bayesian nonparametric priors (and posteriors) are stochastic
processes rather than parametric distributions.  Ferguson
\cite{ferguson:1973} provided the seminal paper for the field of
Bayesian nonparametrics, introducing the Dirichlet process, an
infinite-dimensional generalization of the Dirichlet distribution, now
commonly used as a prior in infinite mixture models.  This is a
popular model (often called the Chinese Restaurant Process) for
classification problems where the number of classes is unknown and to
be inferred from the data.  A formal definition of the Dirichlet
distribution and Dirichlet process can be found in Appendix~B.

Another popular prior in Bayesian nonparametrics is the Gaussian
process prior, which is often used in nonlinear regression contexts.
In fact, one could extend the regression example in the previous
section into the realm of Bayesian nonparametrics by putting a
Gaussian process prior on the function $g$.  Compare this to the
Bayesian parametric counterpart, which puts a prior on the model
parameters $\boldsymbol\beta$.

For further discussion on Bayesian nonparametrics, we refer the reader
to \cite{BNP:2010}.

\subsection{Spectral Density Estimation}

A weakly (or second order) stationary time series $\{X_t\}$ is a
stochastic process that has constant and finite mean and variance over
time, and an autocovariance function $\gamma(h)$ that
depends only on the time lag $h$.  That is, for a zero-mean weakly
stationary process, the autocovariance function has the form
\begin{equation}
  \label{eq:autocovariance}
  \gamma(h) = \mathrm{E}[X_t X_{t+h}], \quad \forall t,
\end{equation}
where $\mathrm{E}[.]$ is the expected value operator, and $t$
represents time.

Assuming an absolutely summable autocovariance function ($\sum_{h =
  -\infty}^{\infty}|\gamma(h)| < \infty$), the (real-valued) spectral
density function $f(\lambda)$ of a zero-mean weakly stationary time series
is defined as
\begin{equation}
  \label{eq:spectral}
  f(\lambda) = \frac{1}{2\pi} \sum_{h = -\infty}^{\infty} \gamma(h)\exp(-\mathrm{i}h\lambda), \quad \lambda \in (-\pi, \pi],
\end{equation}
where $\lambda$ is the angular frequency.  Note that the spectral
density function and autocovariance function are a Fourier transform
pair.  In this paper, we will also call this the \textit{power}
spectral density (PSD) function, although this term is sometimes
reserved for the empirical spectrum (periodogram) in the GW
literature.

For a mean-centered weakly stationary time series $\{X_t\}$ of length
$n$, with spectral density $f(\lambda)$, the Whittle approximation to
the Gaussian likelihood, or simply the \textit{Whittle likelihood}
\cite{whittle:1957} is defined as
\begin{equation}
  \label{eq:whittle}
  L_n(\mathbf{x} | f) \propto \exp\left(-\sum_{l = 1}^{\lfloor u \rfloor} \left(\log f(\lambda_l) + \frac{I_n(\lambda_l)}{f(\lambda_l)} \right) \right),
\end{equation}
where $\lambda_l = 2\pi l/n$ are the positive Fourier frequencies, $u
= (n - 1) / 2$, $\lfloor u \rfloor$ is the greatest integer value less
than or equal to $u$, and $I_n(.)$ is the periodogram defined in
Equation~(\ref{eq:periodogram}).  If the PSD is known, the $\log f$
term in Equation~(\ref{eq:whittle}) is a constant and can be ignored.
The Whittle likelihood has an advantage over the true Gaussian
likelihood as it has a direct dependence on the PSD rather than the
autocovariance function.  The Whittle likelihood is only exact for
Gaussian white noise but works well under certain conditions, even
when the data is not Gaussian \cite{shao:2007}.  More information
about these concepts can be found in any good time series analysis
textbook, such as Brockwell and Davis \cite{brockwell.davis:1991}.

We now need to specify a nonparametric prior for the PSD.  We will
briefly introduce the spectral density estimation technique of
Choudhuri \textit{et al} \cite{choudhuri:2004}, which is based on the
Bernstein polynomial prior of Petrone \cite{petrone:1999a,
  petrone:1999b}.  The Bernstein polynomial prior is a nonparametric
prior for a probability density on [0,~1], and is based on the
Weierstrass approximation theorem that states that any continuous
function on [0,~1] can be uniformly approximated to any desired degree
by a Bernstein polynomial.  If this function is a density on [0,~1],
this Bernstein polynomial is essentially a finite mixture of Beta
densities.  We refer the reader to Appendix~A for a definition of the
Bernstein polynomial and Beta density.  Instead of putting a Dirichlet
prior on the mixture weight vector, the weights are defined via a
probability distribution $G$ on [0,~1] and a Dirichlet process prior
is put on the space of probability distributions on [0,~1].
Appendix~B contains supplementary material on the Dirichlet process.

Since the spectral density is not defined on the unit interval, we
reparameterize $f(\lambda)$, such that
\begin{equation}
  \label{eq:tauq}
  f(\pi\omega) = \tau q(\omega), \quad \omega \in [0, 1],
\end{equation}
where $\tau = \int_0^1 f(\pi\omega) \mathrm{d}\omega$ is the
normalization constant.  To specify a prior on spectral density
$f(\pi\omega)$, we put a Bernstein polynomial prior on $q(\omega)$,
using the following hierarchical scheme:
\begin{itemize}
\item $q(\omega) = \sum_{j = 1}^k
  G\left(\frac{j-1}{k},\frac{j}{k}\right] \beta(\omega | j, k - j +
  1)$, where $G$ is a cumulative distribution function, and $\beta(\omega | a,
  b)$ is a Beta probability density with parameters $a$ and $b$.
\item $G$ is a Dirichlet process distributed random probability
  measure with base measure $G_0$ and precision parameter $M$.
\item $k$ has a discrete probability mass function such that $p(k) \propto
  \exp(-\theta_k k^2), \quad k = 1, 2, \ldots$.
\item $\tau$ has an Inverse-Gamma$(\alpha_{\tau}, \beta_{\tau})$
  distribution.
\item $G$, $k$, and $\tau$ are \textit{a priori} independent.
\end{itemize}

We use the stick-breaking construction of the Dirichlet process by
Sethuraman \cite{sethuraman:1994}, which is an infinite-dimensional
mixture model (defined in Appendix~B).  For computational purposes, we
need to truncate the number of mixture distributions to a large but
finite number $L$.  The choice of a large $L$ will provide a more
accurate approximation but also increase the computation time.  Here,
we choose $L = \max\{20, n^{1/3}\}$.  We therefore reparameterize $G$ to
$(Z_0, Z_1, \ldots, Z_L, V_1, \ldots, V_L)$ such that
\begin{equation}
  \label{eq:sethuraman}
  G = \left(\sum_{l = 1}^L p_l\delta_{Z_l}\right) + \left(1 - \sum_{l = 1}^L p_l\right)\delta_{Z_0},
\end{equation}
where $p_1 = V_1$, $p_l = \left(\prod_{j = 1}^{l - 1}\left(1 -
    V_j\right)\right)V_l$ for $l \geq 2$, $V_l \sim \mathrm{Beta}(1,
M)$ for $l = 1, \ldots, L$, and $Z_l \sim G_0$ for $l = 0, 1, \ldots,
L$. Note that $\delta_{a}$ is a probability density, degenerate at $a$.
That is, $\delta_a = 1$ at $a$ and 0 otherwise.  This yields the
prior mixture of the PSD
\begin{equation}
  \label{eq:betamixture}
  f(\pi\omega) = \tau \sum_{j = 1}^k w_{j, k} \beta(\omega | j, k - j + 1),
\end{equation}
with weights $w_{j, k} = \sum_{l = 0}^L p_l I\{\frac{j - 1}{k} < Z_l
\leq \frac{j}{k}\}$ and $p_0 = 1 - \sum_{l = 1}^L p_l$.

Abbreviating the vector of noise parameters as $\boldsymbol\theta =
(\mathbf{v}, \mathbf{z}, k, \tau)$, the joint prior is therefore
\begin{equation}
  \label{eq:jointprior}
  p(\boldsymbol\theta) \propto \left(\prod_{l = 1}^L M(1 - v_l)^{M-1}\right)\left(\prod_{l = 0}^L g_0(z_l)\right)p(k)p(\tau),
\end{equation}
and is updated using the Whittle likelihood to produce the
unnormalized joint posterior.

This method is implemented as a Metropolis-within-Gibbs MCMC sampler.
In Choudhuri \textit{et al} \cite{choudhuri:2004}, parameters $k$ and
$\tau$ are readily sampled from their full conditional posteriors,
while $\mathbf{V}$ and $\mathbf{Z}$ require the Metropolis algorithm
with Uniform proposals.  Our only variation on this implementation is
our sampling of the smoothness parameter $k$.  We found that a
Metropolis step is faster than sampling from the full conditional.
The original implementation contains a \textsf{for()} loop that
evaluates the log posterior $k_{\max}$ number of times, where
$k_{\max}$ is chosen (during pilot runs) to be large enough to cater
for the roughness of the PSD.  For most well-behaved cases, $k_{\max}
= 50$ will suffice, but the Advanced LIGO PSD requires many more
mixture distributions (by one to two orders of magnitude) due its
steepness at low frequencies.  This is a significant computational
burden, and a well-tuned Metropolis step can therefore outperform the
original implementation.

A discussion of the Dirichlet process and stick-breaking
representation can be found in Appendix~B.

\subsection{Signal Reconstruction}

To reconstruct a rotating core collapse GW signal that is embedded in
noise, we use the (parametric) principal component regression (PCR)
method described in \cite{heng:2009, roever:2009, edwards:2014}.  That
is, let
\begin{equation}
  \label{eq:pcr}
  \tilde{\mathbf{y}} = \tilde{\mathbf{X}}\boldsymbol\beta + \tilde{\boldsymbol\epsilon},
\end{equation}
where $\tilde{\mathbf{y}}$ is the length $n$ frequency-domain GW data
vector, $\tilde{\mathbf{X}}$ is the $n \times d$ matrix of the $d$
frequency-domain principal component basis vectors, $\boldsymbol\beta$
is the vector of signal reconstruction parameters (PC coefficients),
and $\tilde{\boldsymbol\epsilon}$ is the frequency-domain noise vector
with a \textit{known} PSD.  We assume flat priors on
$\boldsymbol\beta$.  It is important to highlight that useful
astrophysical information (such as the ratio of kinetic to
gravitational potential energy of the inner core at bounce, and
precollapse differential rotation) can be extracted by regressing the
posterior means of the PC coefficients $\boldsymbol\beta$ on the known
astrophysical parameters from the waveform catalogue, and sampling
from the posterior predictive distribution \cite{edwards:2014}.

We include an additional Gibbs step in the MCMC sampler described in
the previous section to simultaneously reconstruct a rotating core
collapse GW signal, whilst also estimating the noise power spectrum.
Omitting the conditioning on the data for clarity, we sequentially
sample the full set of conditional posterior densities
$p(\boldsymbol\theta | \boldsymbol\beta)$ and $p(\boldsymbol\beta |
\boldsymbol\theta)$, where $\boldsymbol\theta = (\mathbf{v},
\mathbf{z}, k, \tau)$ are the noise parameters defined in the previous
section, and $\boldsymbol\beta$ are the signal reconstruction
parameters.  That is, we sample in a cycle from the full conditional
posterior distribution of the signal parameters, given the PSD
parameters, and the full conditionals of the PSD parameters, given the
signal parameters.  This set-up is called a \textit{blocked}
Gibbs sampler.

To sample the signal parameters, we fix the most recent MCMC sample of
the PSD parameters.  The conditional posterior of $\boldsymbol\beta$
is
 \begin{equation}
   \label{eq:signalPosterior}
   \mathrm{P}(\boldsymbol\beta | \boldsymbol\theta) = \mathrm{N}(\boldsymbol\mu, \boldsymbol\Sigma) 
 \end{equation}
 where $\boldsymbol\Sigma =
 (\tilde{\mathbf{X}}^{'}\mathbf{D}^{-1}\tilde{\mathbf{X}})^{-1}$ and
 $\boldsymbol\mu = \boldsymbol\Sigma
 \tilde{\mathbf{X}}^{'}\mathbf{D}^{-1}\tilde{\mathbf{y}}$.  Here
 $\mathbf{D} = 2\pi \times
 \mathrm{diag}\left(f(\boldsymbol\lambda)\right)$ is the noise
 covariance matrix, and $f(\boldsymbol\lambda)$ is the most recent
 estimate of the PSD.  More explicitly, at iteration $i + 1$ in the
 blocked Gibbs sampling algorithm:
\begin{enumerate}
\item Create time-domain noise vector: $\boldsymbol\epsilon^{(i + 1)}
  = \mathbf{y} - \mathbf{X}\boldsymbol\beta^{(i)}$.  Due to the
  linearity of the Fourier transform, $\boldsymbol\beta$ will be the
  same no matter if we are in the time-domain or frequency-domain.
\item Sample the PSD parameters $\boldsymbol\theta^{(i + 1)} |
  \boldsymbol\beta^{(i)}$ using the method of Section~II~C.
\item Sample the signal parameters $\boldsymbol\beta^{(i + 1)} |
  \boldsymbol\theta^{(i + 1)}$ using
  Equation~(\ref{eq:signalPosterior}) (since the PSD in iteration $i +
  1$ is now known).
\end{enumerate}

\subsection{Non-stationary Noise}

As mentioned in Section~II~C, stationary noise has a constant and
finite mean and variance over time, and an autocovariance function
that depends only on the time lag.  Non-stationary noise does not meet
these requirements, and has a time-varying spectrum.  Stationarity of
a time series can be tested using classical hypothesis tests such as
the Augmented Dickey-Fuller test \cite{adf:1984}, Phillips-Perron unit
root test \cite{phillips:1988}, and Kwiatwoski-Phillips-Schmidt-Shin
(KPSS) test \cite{kpss:1986}.

To accommodate non-stationary noise, we adapt an idea presented by
Rosen \textit{et al} \cite{rosen:2012}, that assumes a time series can
be broken down into locally stationary segments.  In their paper, they
treat the number of stationary components of a non-stationary time
series as unknown, and use RJMCMC \cite{green:1995} to estimate the
segment breaks.

In a similar fashion, we break a non-stationary time series (or GW
data stream) into $J$ equal segments.  We have two requirements for
the length of these segments: the segment length is large enough for
the Whittle approximation to be valid; and the segments are locally
stationary according to heuristics or formal stationarity hypothesis
tests.  This approach fits nicely into our current MCMC framework.
For each segment, we estimate the PSD using the nonparametric method
introduced in Section~II~C.  A benefit of this approach is that
change-points in the PSD can be detected without using RJMCMC.

The conditional posterior density for all noise model parameters
$\boldsymbol\theta$ is the following product
\begin{equation}
  \label{eq:productPosterior}
  \pi(\boldsymbol\theta | \boldsymbol\beta, \tilde{\mathbf{y}}) = \prod_{j = 1}^J \pi_j(\boldsymbol\theta_j | \boldsymbol\beta, \tilde{\mathbf{y}}_j) ,
\end{equation}
where $\pi_j(\boldsymbol\theta_j | \boldsymbol\beta,
\tilde{\mathbf{y}}_j)$ is the conditional posterior density of the
model parameters $\boldsymbol\theta_j$ in the $j^{\mathrm{th}}$
segment given the signal parameters $\boldsymbol\beta$ and the
$j^{\mathrm{th}}$ segment of data $\tilde{\mathbf{y}}_j$.

Note that under this set-up, the PC coefficients $\boldsymbol\beta$ do
not depend on segments $j = 1, 2, \ldots, J$, since we require one set
of PC coefficients (not $J$ sets) to reconstruct a rotating core
collapse GW signal.  

To sample $\boldsymbol\beta | \boldsymbol\theta$, we use the same
approach presented in Section~II~D.  The only difference is in the
construction of the noise covariance matrix.  This is constructed as
$\mathbf{D} = 2\pi \times \mathrm{diag}(f_1(\boldsymbol\lambda),
f_2(\boldsymbol\lambda), \ldots, f_J(\boldsymbol\lambda))$, where
$f_j(\boldsymbol\lambda)$ is the PSD of the $j^{\mathrm{th}}$ noise
segment.

\section{Results}

For the following examples, we set $L = \max\{20, n^{1/3}\}$, and use
the non-informative prior set-up of Choudhuri \textit{et al}
\cite{choudhuri:2004}.  That is, let $G_0 \sim \mathrm{Uniform}[0, 1],
M = 1, \alpha_{\tau} = \beta_{\tau} = 0.001$, and $ \theta_k = 0.01$.
We use $k_{\max} = 50$ for most examples, and $k_{\max} = 400$ for the
example with simulated Advanced LIGO noise to cater for the steep drop
in the PSD at low frequencies.

For the examples with a signal embedded in noise, we use a
$\mathrm{Uniform}(-\infty, \infty)$ prior on the signal reconstruction
parameters $\boldsymbol\beta$, and let $d = 25$ PCs.  For a discussion
on the optimal choice of PCs, we refer the reader to
\cite{edwards:2014}.  We also scale the signals to a signal-to-noise
ratio (SNR) of $\varrho = 50$.  Here SNR (for $n$ even) is defined as
\begin{equation}
  \label{eq:SNR}
  \varrho = \sqrt{2 \sum_{j=0}^{n / 2 + 1} \frac{|\tilde{s}(\lambda_j)|^2}{|\tilde{\epsilon}(\lambda_j)|^2}} ,
\end{equation}
where $\lambda_j$ are the positive
Fourier frequencies, $\tilde{s}(.)$ is the Fourier transformed
signal, and $\tilde{\epsilon}(.)$ is the Fourier transformed noise
series.  Note that for the zero and Nyquist frequencies, the factor of
2 in Equation~(\ref{eq:SNR}) becomes a factor of 4.

The value of $\varrho = 50$ is physically motivated, as we would
expect to see an SNR of approximately 50 to 170 for rotating core
collapse supernova GWs at a distance of $10~\mathrm{kpc}$.  We
therefore demonstrate how the method works for the lower end of this
range.

The units for frequency in most examples are radians per second
($\mathrm{rad/s}$).  In the example using simulated Advanced LIGO
noise, we rescale to kilohertz ($\mathrm{kHz}$).  PSD units are the
inverse of the frequency units, and the PSD figures are scaled
logarithmically.  GW strain amplitude is unitless.

For all examples, we run the MCMC sampler for $150,000$ iterations,
with a burn-in period of $50,000$ and thinning factor of $10$.  This
results in $10,000$ samples retained.

\subsection{Estimating the PSD of Non-Gaussian Coloured Noise}

To demonstrate how our model is capable of dealing with non-Gaussian
transients in the data (or \textit{glitches} as they are sometimes
called in GW data analysis), we provide an illustrative toy example,
using coloured noise generated from a first order autoregressive
process, abbreviated to AR(1).

A mean-centered AR(1) process $\{X_t\}$ is defined as
\begin{equation}
  \label{eq:AR1}
  X_t = \rho X_{t-1} + \epsilon_t, \quad t = 1, 2, \ldots, n,
\end{equation}
where $\rho$ is the first order autocorrelation, and $\epsilon_t$ is a
white noise process (not necessarily Gaussian), with zero mean and
constant variance $\sigma^2_{\epsilon}$.  With this formulation, we
see how the current observation at time $t$ depends on the previous
observation at time $t-1$ through $\rho$, as well as some white noise
$\epsilon_t$, often referred to as \textit{innovations} or the
\textit{innovation process} in time series literature. 

The AR(1) model is a useful example here since it has a well-defined
theoretical spectral density that we can compare our results against.
Assuming $|\rho| < 1$, the AR(1) process is stationary and has
spectral density
\begin{equation}
  \label{eq:AR1psd}
  f(\lambda) = \frac{\sigma^2_{\epsilon}}{1 + \rho^2 - 2 \rho \cos 2\pi \lambda}, \quad \lambda \in (-\pi, \pi].
\end{equation}

As seen in Equation~(\ref{eq:AR1psd}), the AR(1) process has a PSD
that is not flat, and the noise in our toy example is coloured
(non-white), with correlations between frequencies --- typical of what
we would expect with real Advanced LIGO noise.  As the AR(1) process
has a coloured spectrum, and white noise has a flat spectrum, we will
call use the term \textit{innovations} to refer to the white noise
component of the model to avoid confusion.

For our example, rather than using Gaussian innovations, which is the
most common innovation process used in autoregressive models, we use
Student-$t$ innovations with $\nu = 3$ degrees of freedom.  The choice
of $\nu = 3$ degrees of freedom is the smallest integer that results
in a Student-$t$ model with finite variance (a requirement for the
innovation process $\{\epsilon_t\}$ of an AR(1) model).  This model
has wider tails than that of the Gaussian model (and in fact the
widest tails possible whilst maintaining the finite variance
requirement), meaning we can expect extreme values in the tails of the
distribution to occur more often.  This will be our proxy for
glitches.

We refer the reader to a relevant time series analysis textbook such as
Brockwell and Davis \cite{brockwell.davis:1991} for further
information on AR(1) processes.

For this example, we generate a length $n = 2^{12}$ AR(1) process with
$\rho = -0.9$ and Student-$t$ innovations with $\nu = 3$ degrees of
freedom.  Let this (stationary) time series have sampling interval
$\Delta_t = 1 / 2^{14}$ (the same as Advanced LIGO).  The data set-up
can be seen in Figure~\ref{fig:ar1st.plot}.

\begin{center}
\includegraphics[width=1\linewidth]{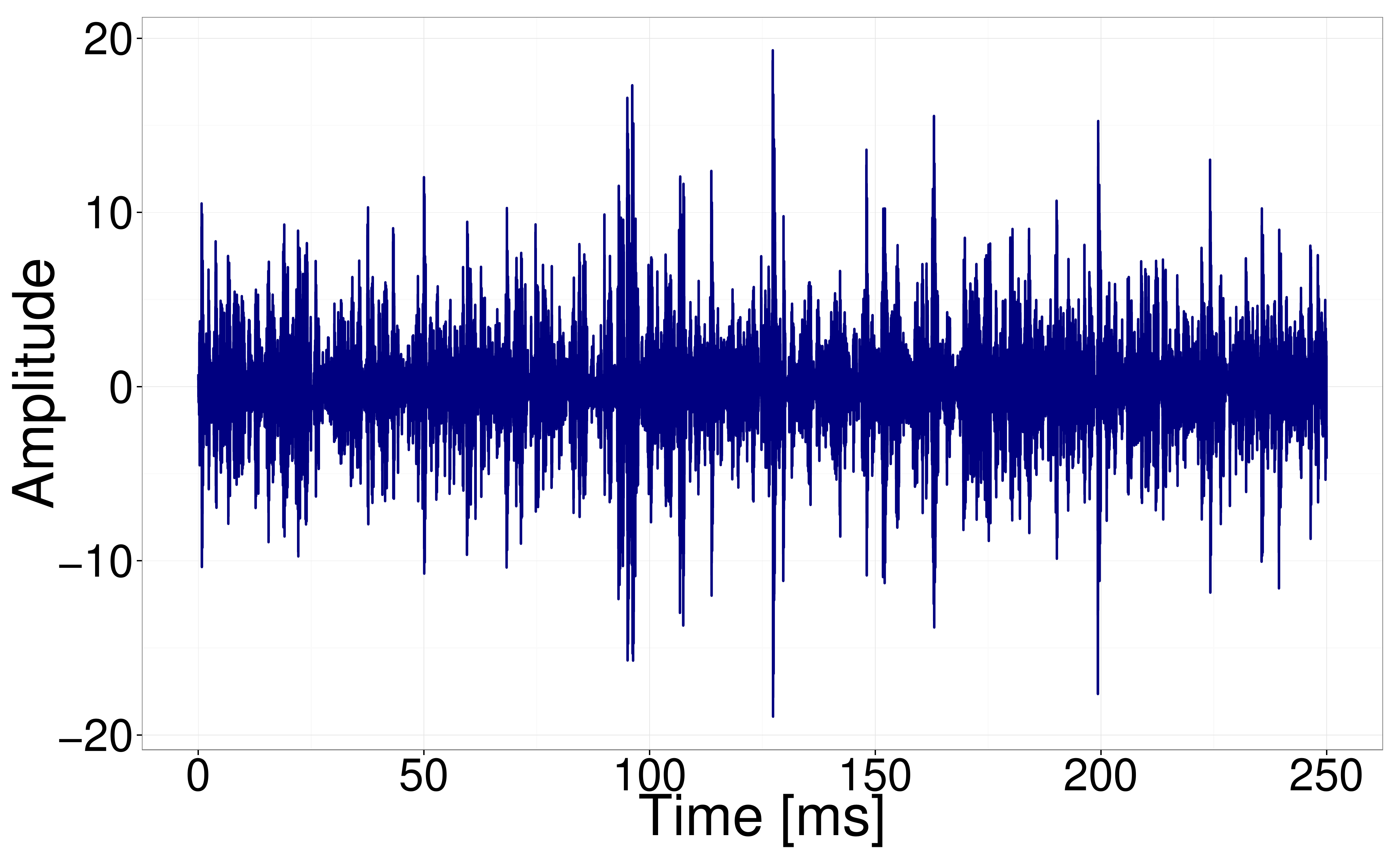}
\captionof{figure}{Simulated stationary AR(1) process with first-order
  autocorrelation $\rho = -0.9$, and Student-$t$ innovations ($\nu =
  3$ degrees of freedom).}
\label{fig:ar1st.plot}
\end{center}

We can see the effect of using $\nu = 3$ degrees of freedom in
Figure~\ref{fig:ar1st.plot}.  Notice how there are transient high
amplitude non-Gaussian events.  These are a result of the wide-tailed
nature of the Student-$t$ density.  It would be very unlikely to see
these high amplitude events if the innovation process was Gaussian.

We now run the noise-only algorithm of Section~II~C to demonstrate that we
can accurately characterize a non-Gaussian noise PSD.

\begin{center}
\includegraphics[width=1\linewidth]{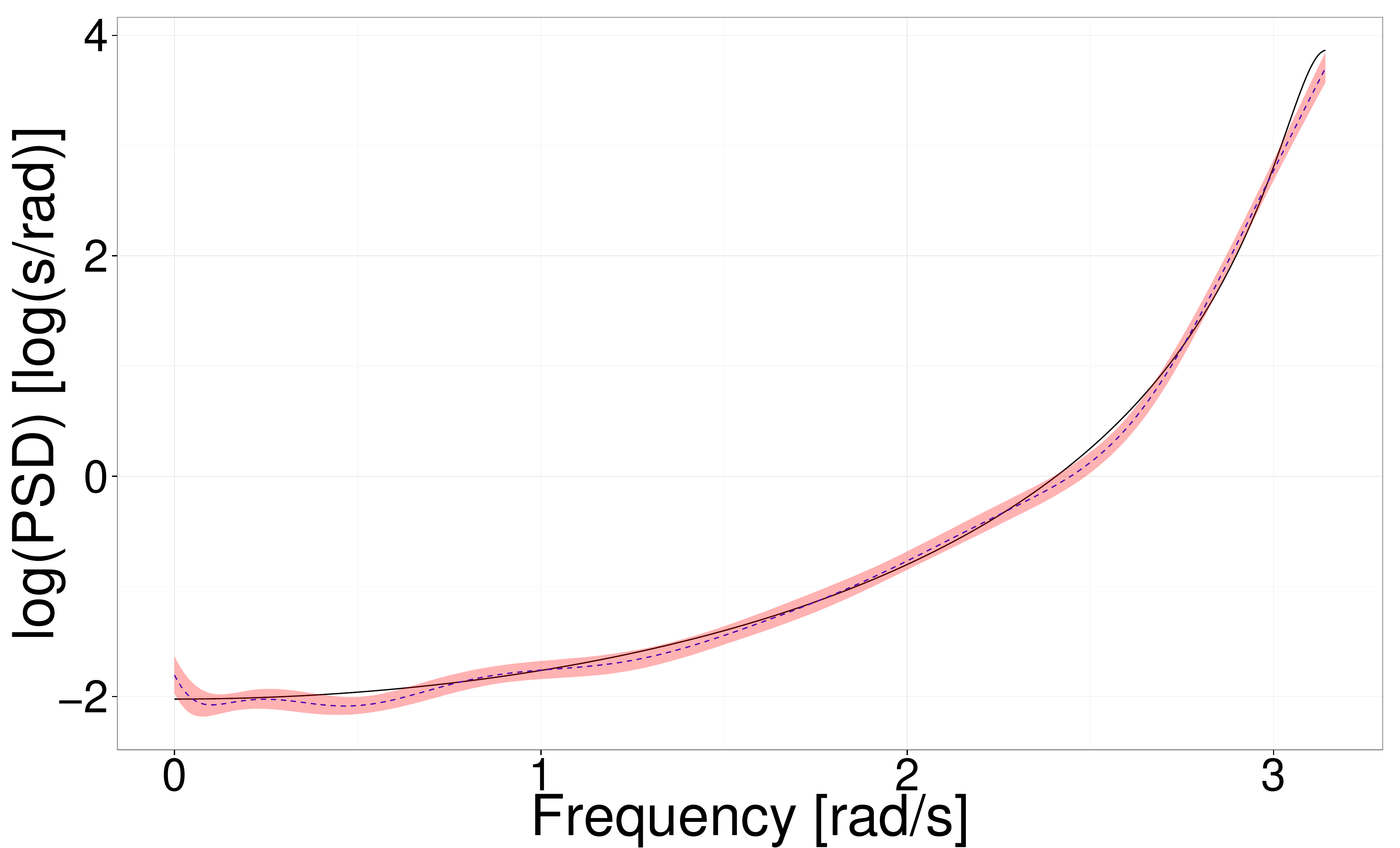}
\captionof{figure}{Estimated log PSD of the AR(1) time series in
  Figure~\ref{fig:ar1st.plot}.  90\% credible region (shaded pink) and
  posterior median log PSD (dashed blue), superimposed with the true
  log PSD (solid black).}
\label{fig:ar1stPSD}
\end{center}

The estimated point-wise posterior median log PSD in
Figure~\ref{fig:ar1stPSD} is very close to the true log PSD, and the
90\% credible region generally contains the true log PSD.  This
demonstrates that even if there are non-Gaussian transients in the
data (which is certainly the case for real LIGO data), this PSD
estimation method performs well.  This is however not surprising as
the Whittle likelihood gives a good approximation to Gaussian
and some non-Gaussian likelihoods \cite{shao:2007}.

\subsection{Extracting a Rotating Core Collapse Signal in Stationary Coloured Noise}

In this example, we aim to extract a rotating GW signal from noisy
data using the blocked Gibbs sampler described in Section~II~D.  We
embed the $A1O12.25$ rotating core collapse GW signal from the
Abdikamalov \textit{et al} \cite{abdikamalov:2014} test catalogue
(i.e., a signal not part of the base catalogue used to create
the PC basis functions) in AR(1) noise with $\rho = 0.9$.  For
clarity, let this process have a Gaussian white noise innovation
process with $\sigma_{\epsilon}^2 = 1$.  Let the time series be length
$n = 2^{12}$, which corresponds to $1/4~\mathrm{s}$ of data at the
Advanced LIGO sampling rate.  The signal is scaled to have a SNR of
$\varrho = 50$.  The reconstructed signal can be seen in
Figure~\ref{fig:recon.stationary}.

The rotating core collapse GW signal in
Figure~\ref{fig:recon.stationary} is reconstructed particularly well
during the collapse and bounce phases (the first few peaks/troughs).
The post-bounce ring-down oscillations are usually poorly estimated
due to stochastic dynamics \cite{abdikamalov:2014, edwards:2014}, but
are acceptable for this particular example.

\begin{center}
\includegraphics[width=1\linewidth]{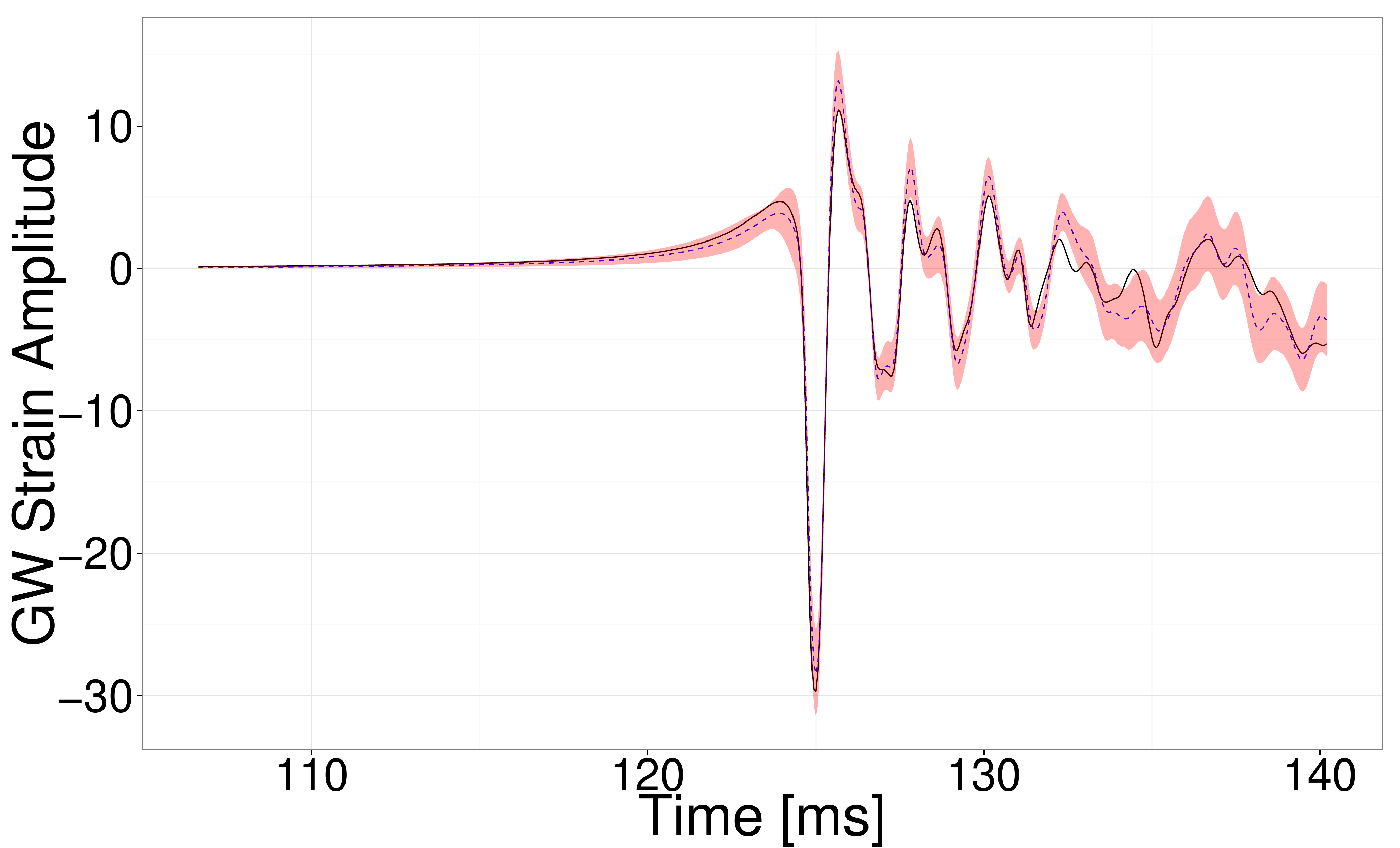}
\captionof{figure}{Reconstructed rotating core collapse GW signal.
  90\% credible region (shaded pink) and posterior median signal (dashed
  blue), superimposed with true $A1O12.25$ GW signal from the
  Abdikamalov \textit{et al} \cite{abdikamalov:2014} test catalogue
  (solid black).}
\label{fig:recon.stationary}
\end{center}

In this example, the signal parameters were simultaneously estimated
with the noise PSD using the blocked Gibbs sampler described in
Section~II~D.  We now compare the performance of the estimated noise
PSD with and without a signal present.  That is, we compare the noise
PSD estimates between the algorithms presented in Section~II~C
(noise-only model) and Section~II~D (signal-plus-noise model), using
the same noise series for both models.

\begin{center}
\includegraphics[width=1\linewidth]{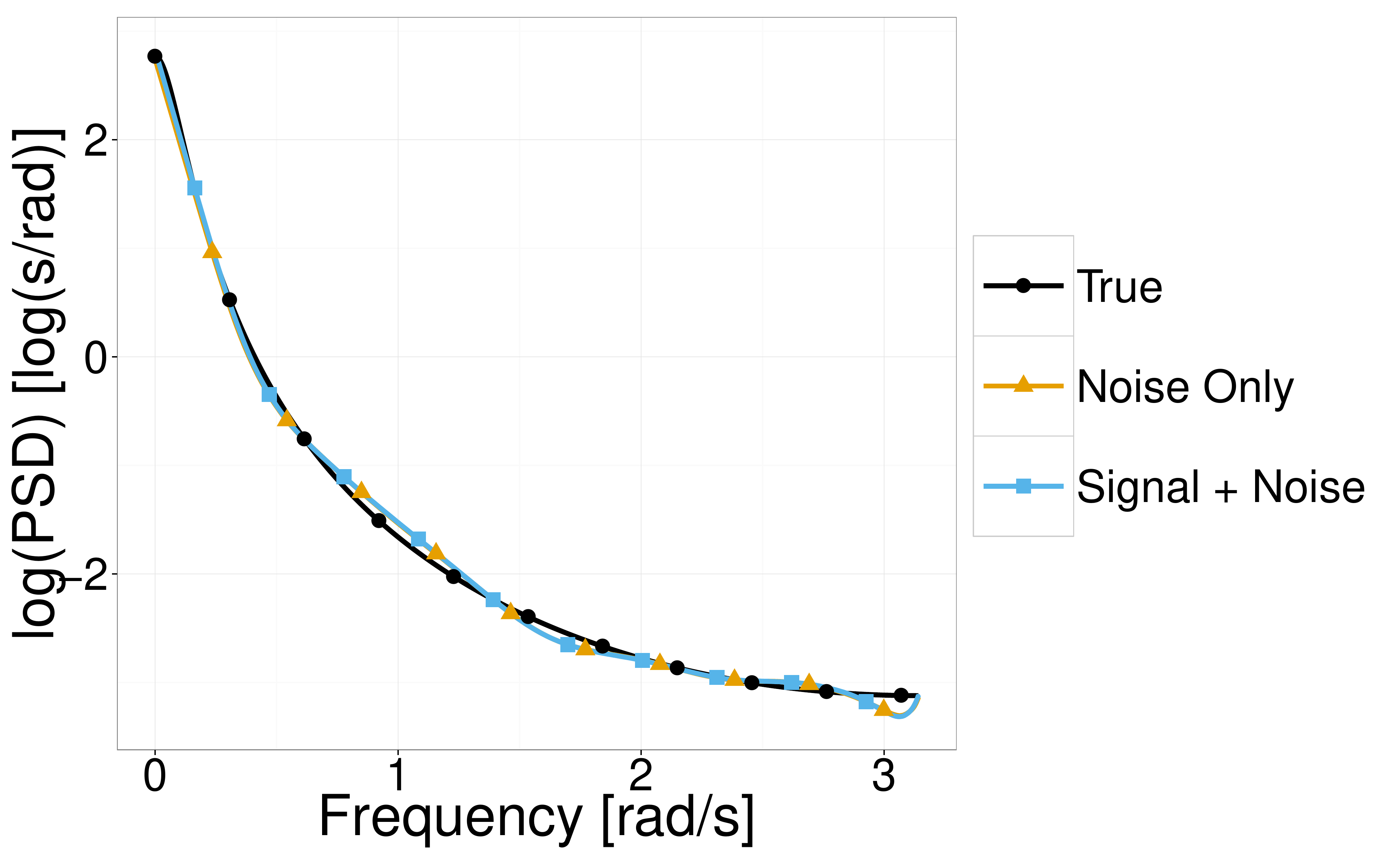}
\captionof{figure}{Comparison of the noise PSD estimates for the
  noise-only and signal-plus-noise models.  Plotted are the point-wise
  posterior median log noise PSDs with and without a GW signal.  The
  true log PSD of the AR(1) noise series is overlaid.}
\label{fig:comparePSD}
\end{center}

We can see in Figure~\ref{fig:comparePSD} that both models (noise-only
and signal-plus-noise) perform similarly when estimating the PSD of
coloured Gaussian noise.  The posterior median log PSDs are
approximately equal, and are very close to the true log PSD of an
AR(1) process with $\rho = 0.9$.  This is a useful robustness check,
and demonstrates that we are successfully decoupling the signal from
the noise.

\subsection{Comparing Input and Reconstruction Parameters}

As there is no analytic form linking the astrophysical parameters of a
rotating core collapse stellar event to its GW signal, we can only
approximate the GW signal using statistical methods.  We do this using
PCR, but this means that there are no true input parameters that we
can compare with the estimated signal reconstruction parameters.
However, if one were to create a fictitious signal as a known linear
combination of PCs, we could demonstrate the algorithm's performance in
estimating the signal reconstruction parameters.

Consider the following fictitious rotating core collapse GW signal
\begin{equation}
  \label{eq:linearcomb}
  \mathbf{y} = \sum_{i = 1}^d \alpha_i \mathbf{x_i},
\end{equation}
where $\mathbf{y}$ is the length $n$ signal, $(\mathbf{x_1},
\mathbf{x_2}, \ldots, \mathbf{x_d})$ are the $d$ PC basis vectors of
length $n$, and $(\alpha_1, \alpha_2, \ldots, \alpha_d)$ are the
``true'' weights, or PC coefficients.  To randomize the weights, we
randomly sample each from the standard normal distribution.

In this example, we embed the fictitious length $n = 2^{12}$ GW signal
in AR(1) noise with $\rho = 0.9$ and Gaussian innovations with
$\sigma_{\epsilon} = 1$.  We set $d = 10$.  We rescale the signal
to have SNR $\varrho = 50$, and after the algorithm has run, we
rescale our estimated PC coefficients back to the original level for
comparison. 

\begin{center}
\includegraphics[width=1\linewidth]{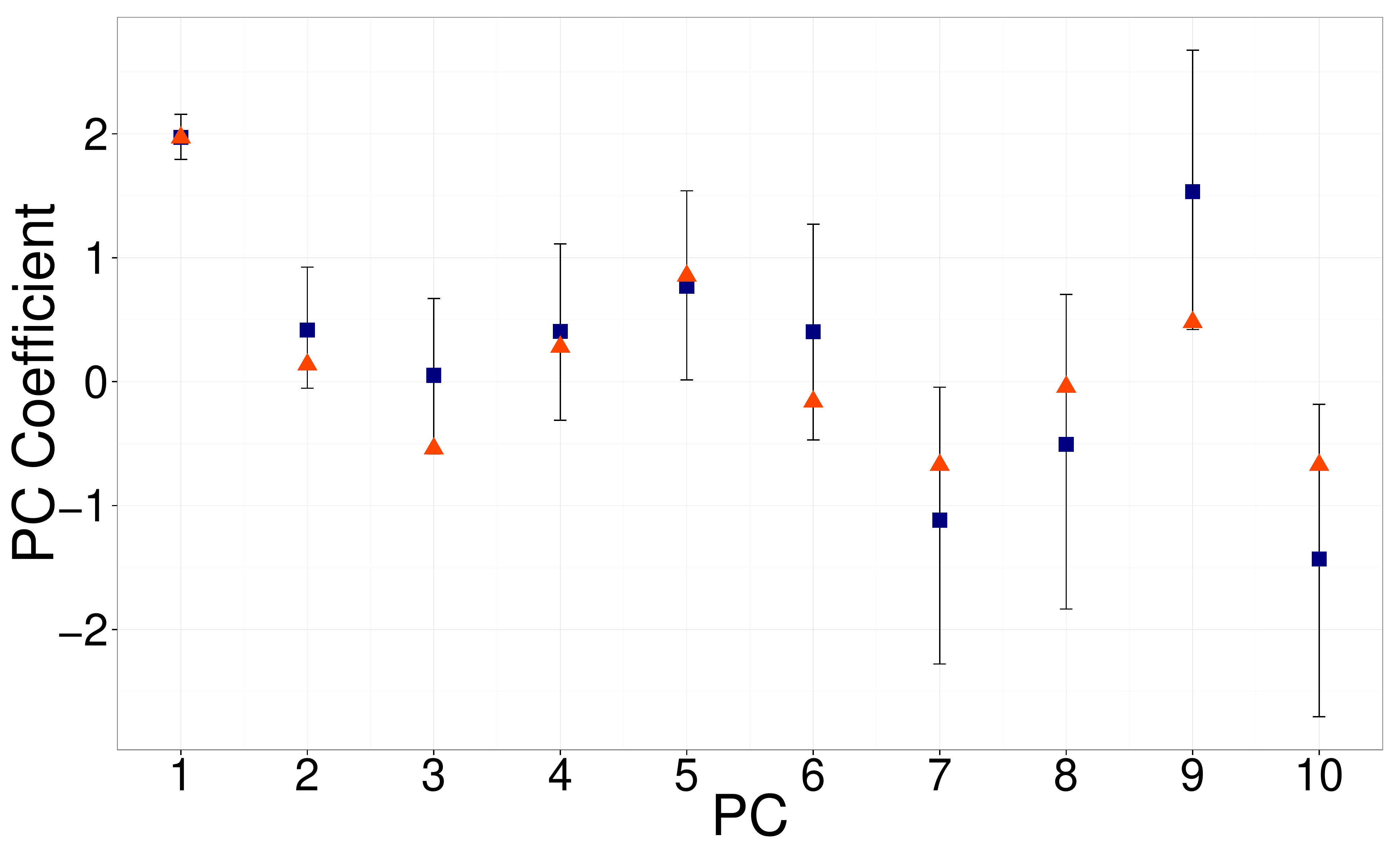}
\captionof{figure}{Posterior median PC coefficients (blue square) and
  ``true'' PC coefficients (orange triangle) for the 10 PCs of a
  fictitious GW signal embedded in AR(1) noise.  The error bands are
  the 95\% credible intervals.}
\label{fig:comparePCs}
\end{center}

It can be seen in Figure~\ref{fig:comparePCs} that the ``true'' PC
coefficients are generally contained within the 95\% credible
intervals, demonstrating that the algorithm can estimate a signal's
input parameters well in the presence of stationary coloured noise.
Notice also that the credible intervals widen as the principal
component number increases.  This is due to the fact that higher
numbered PCs explain lower amounts of variation in the waveform
catalogue, resulting in lower amplitude waves.  We would therefore be
more uncertain about these PCs embedded in noise.

\subsection{Extracting a Rotating Core Collapse Signal in Time-Varying Coloured Noise}

Non-stationary noise has a time-varying spectrum.  To illustrate how
our method can handle non-stationarities (or change-points in the
spectral structure), we simulate a noise series with $J = 2$ locally
stationary components of equal length $n_1 = n_2 = 2^{12}$.  The first
segment of the noise series is generated from an AR(1) process with
$\rho = 0.5$.  The second noise segment comes from an AR(1) process
with $\rho = -0.75$.  Both segments use a Gaussian innovation process
with variance $\sigma_{\epsilon}^2 = 1$ for clarity.  We embed part of
the $A1O8.25$ waveform from the Abdikamalov \textit{et al} catalogue
\cite{abdikamalov:2014}.  This waveform is in the test set, not
included in the construction of PC basis functions.  The data set-up
can be seen in Figure~\ref{fig:signalNoise.nstat}.

\begin{center}
\includegraphics[width=1\linewidth]{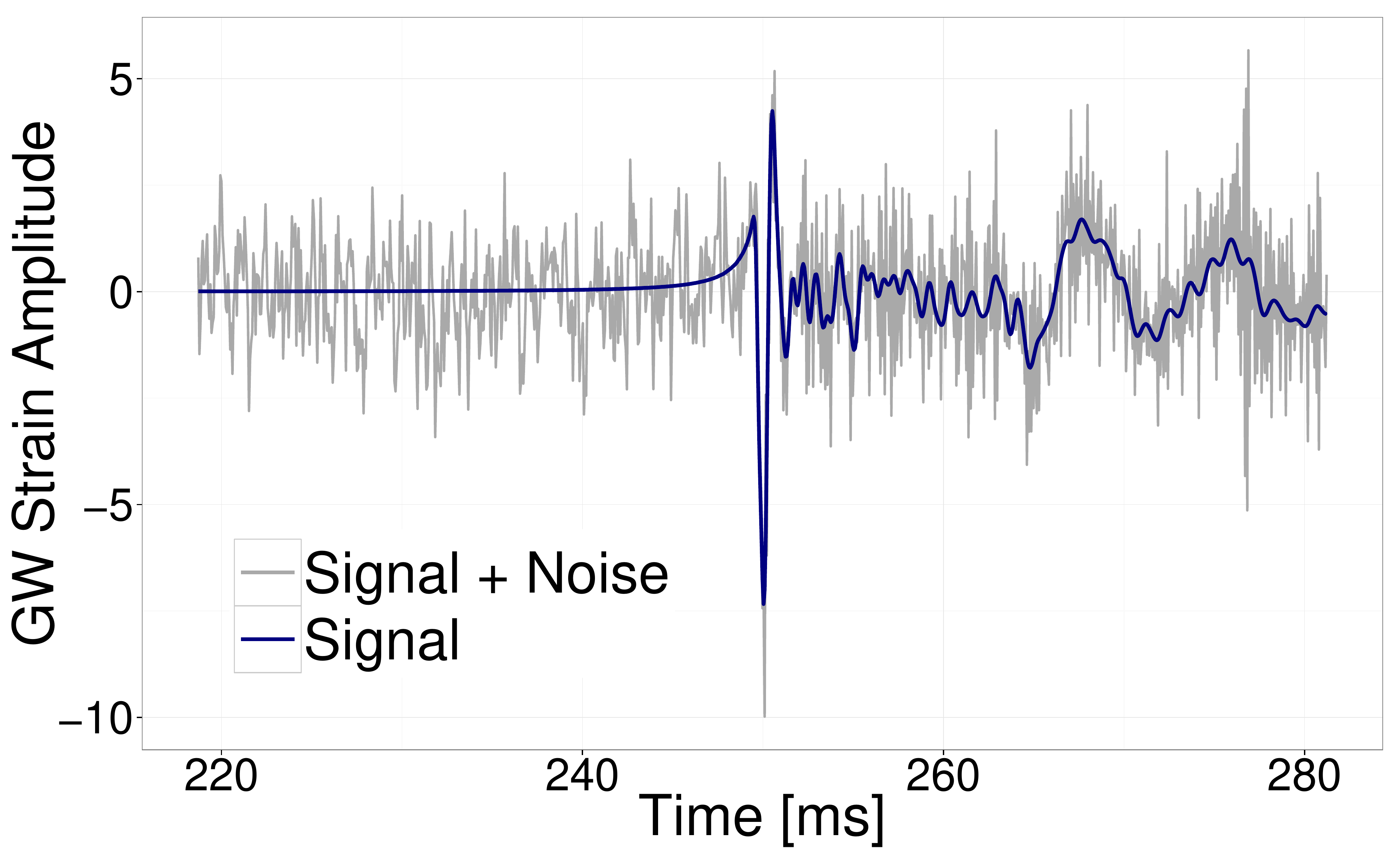}
\captionof{figure}{Snapshot of the signal superimposed on signal plus
  noise.  The noise series has length $n = n_1 + n_2 = 2^{13}$ and is
  segmented into two equal parts.  The first half of the noise is
  generated from an AR(1) with $\rho = 0.5$, and the second half is
  generated from an AR(1) with $\rho = -0.75$.  Both segments use a
  Gaussian innovation process with variance $\sigma_{\epsilon}^2 = 1$.
  The $A1O8.25$ rotating core collapse GW signal from the Abdikamalov
  \textit{et al} test catalogue \cite{abdikamalov:2014} is embedded in
  this noise with a SNR of $\varrho = 50$.}
\label{fig:signalNoise.nstat}
\end{center}

The aim here is to simultaneously estimate both noise PSDs, as well as
reconstructing the embedded GW signal using the method described in
Section~II~E.  Here we are assuming the change-point between the two
noise series is known, though we will demonstrate in the next section
that our method can locate unknown change-points.  

Notice the difference between the first half of the noise series
compared with the second half.  Each segment has a different
dependence structure, and are therefore coloured differently in the
frequency-domain.  This results in a different time-domain morphology.
Estimates of the noise PSDs can be seen in
Figures~\ref{fig:psd1.signal.nstat}~and~\ref{fig:psd2.signal.nstat}.

\begin{center}
\includegraphics[width=1\linewidth]{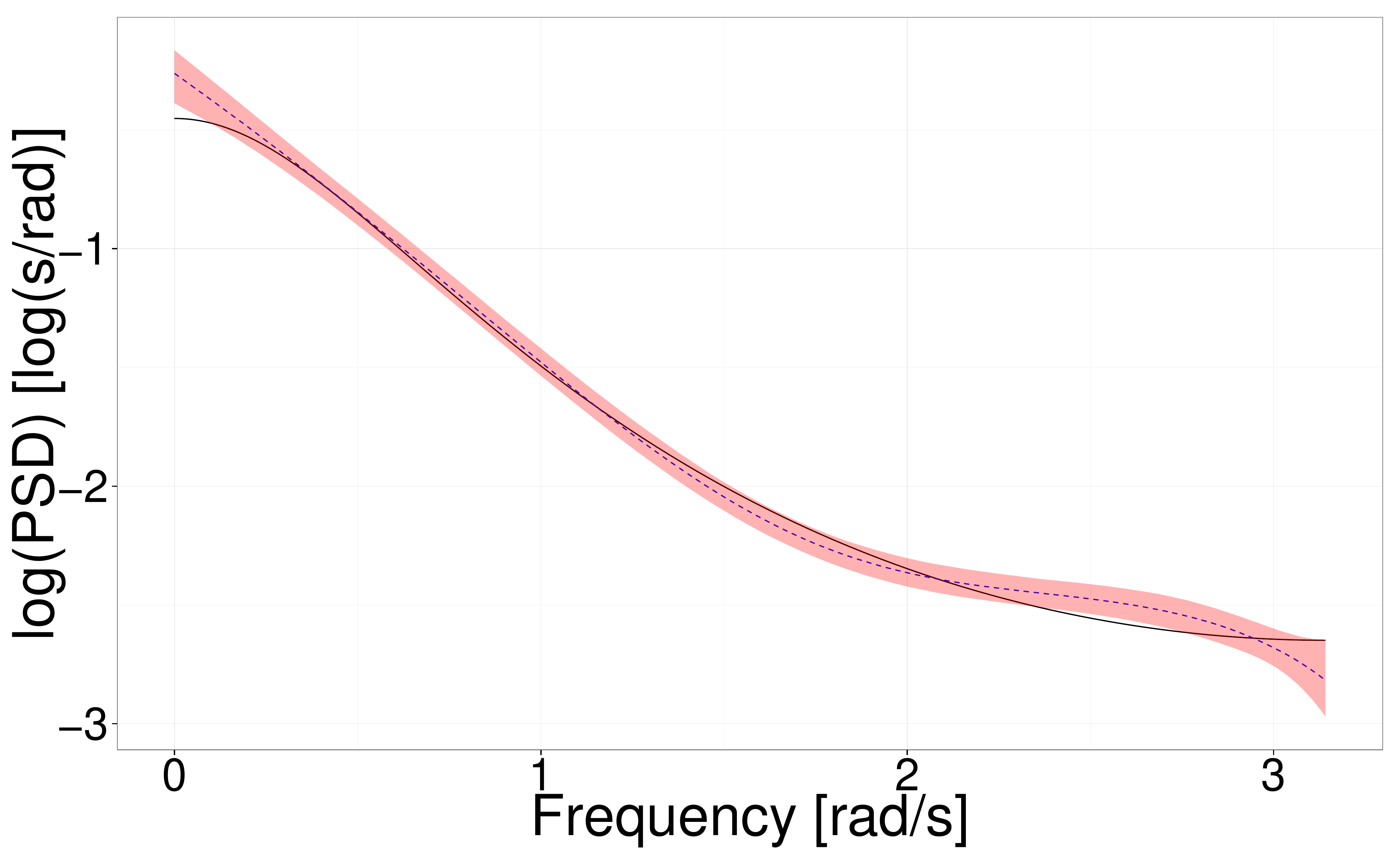}
\captionof{figure}{Spectral density estimate of the first noise
  segment ($\rho = 0.5$) from Figure~\ref{fig:signalNoise.nstat}. 90\%
  credible region (shaded pink), posterior median log PSD (dashed blue), and
  theoretical log PSD (solid black).}
\label{fig:psd1.signal.nstat}
\end{center}

\begin{center}
\includegraphics[width=1\linewidth]{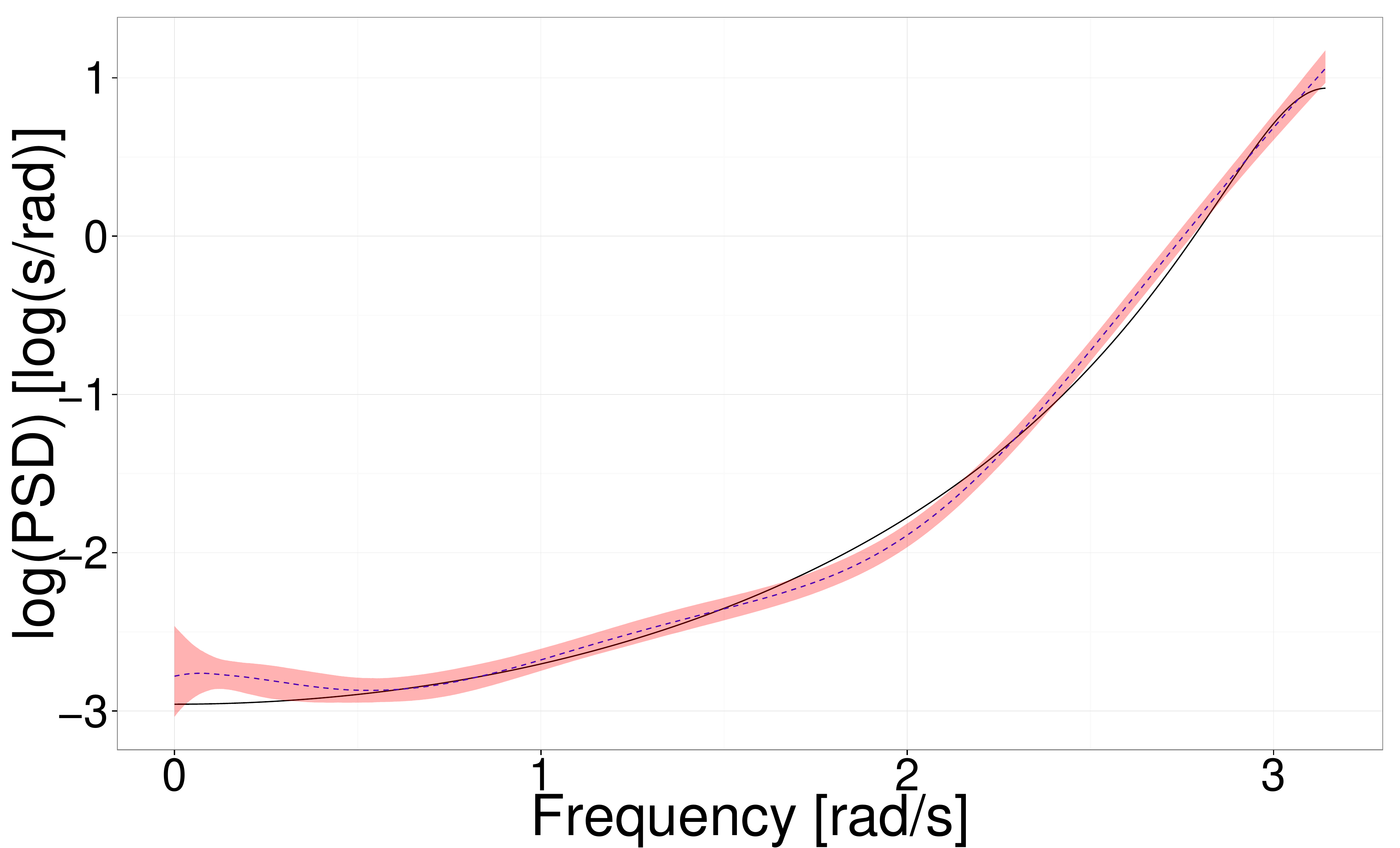}
\captionof{figure}{Spectral density estimate of the second noise
  segment ($\rho = -0.75$) from Figure~\ref{fig:signalNoise.nstat}. 90\%
  credible region (shaded pink), posterior median log PSD (dashed blue), and
  theoretical log PSD (solid black).}
\label{fig:psd2.signal.nstat}
\end{center}

Figures~\ref{fig:psd1.signal.nstat}~and~\ref{fig:psd2.signal.nstat}
show the estimated log PSDs for the two noise segments. The point-wise
posterior median log PSDs are close to the true log PSDs, and the 90\%
credible regions for both segments mostly contain the true log PSDs,
but veer slightly off towards the low frequencies.  Due to posterior
consistency of the PSD, these estimates will only get better as the
sample size increases.  Slight imperfections in the PSD estimates may
not be such a problem if the embedded GW signal is extracted well,
which happens to be the case in this example.  The extracted signal
can be seen in Figure~\ref{fig:recon.nstat}.

\begin{center}
\includegraphics[width=1\linewidth]{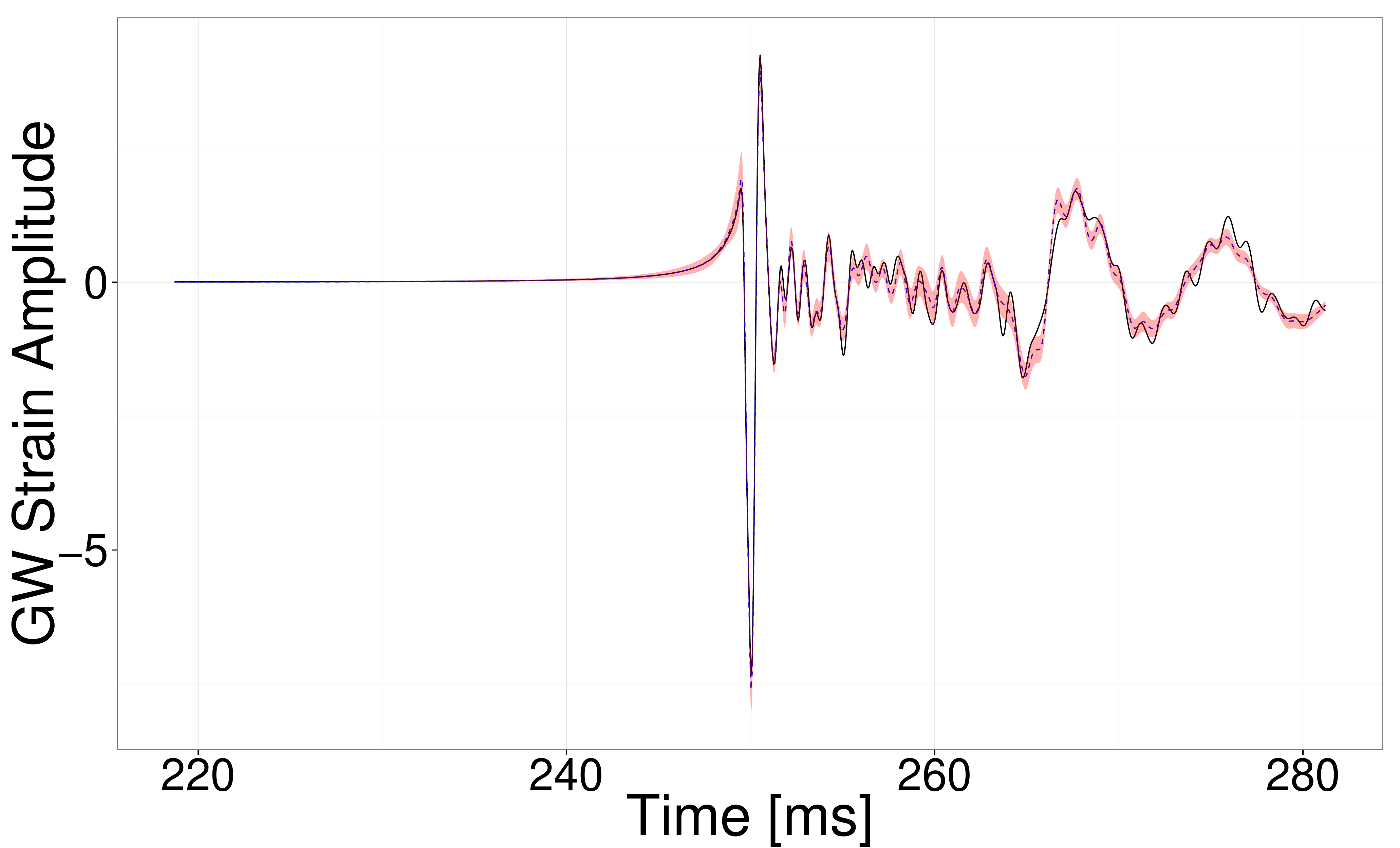}
\captionof{figure}{Reconstructed rotating core collapse GW.  90\%
  credible region (shaded pink) and posterior median signal (dashed
  blue), superimposed with true $A1O8.25$ GW signal from the
  Abdikamalov \textit{et al} \cite{abdikamalov:2014} test catalogue
  (solid black).  The first half of the signal was embedded in AR(1)
  noise with $\rho = 0.5$, and the second half had AR(1) noise with
  $\rho = -0.75$.  Both noise segments had Gaussian white noise with
  $\sigma_{\epsilon}^2 = 1$.}
\label{fig:recon.nstat}
\end{center}

The 90\% credible region for the reconstructed GW signal in
Figure~\ref{fig:recon.nstat} generally contains the true signal, and
has performed particularly well during collapse and bounce.  Again,
the post-bounce ring-down oscillations usually have the poorest
reconstruction through the time series, but has performed remarkably
well in this example, regardless of the slight imperfections of the
PSD estimates.

\subsection{Detecting a Spectral Change-Point}

Consider a change-point problem similar to that of the previous
section, where a time series exhibits a change in its spectral
structure somewhere in the series.  A valuable consequence of the
algorithm presented in Section~II~E is its ability to detect
change-points regardless of whether the change-point occurs within a
segment or on the boundary.  For the following examples, let $n =
2^{12}$ and break this into $J = 32$ equal length segments.  For
clarity, assume the time series does not contain an embedded GW
signal.

First consider the case where the change-point occurs on the boundary
of two noise series.  Let $n_1 = n_2 = 2^{11}$ be the lengths of each
noise series, and let the first half of the time series be generated
from an AR(1) with $\rho = 0.5$, and the second half from an AR(1)
with $\rho = -0.75$.  Both AR(1) processes have additive Gaussian
innovations with $\sigma_{\epsilon}^2 = 1$.  In this example, the
change-point occurs exactly halfway through the series.
Figure~\ref{fig:spectroAR1_boundary} shows a time-frequency map of the
estimated log PSDs for each segment.

\begin{center}
 \includegraphics[width = 1\linewidth]{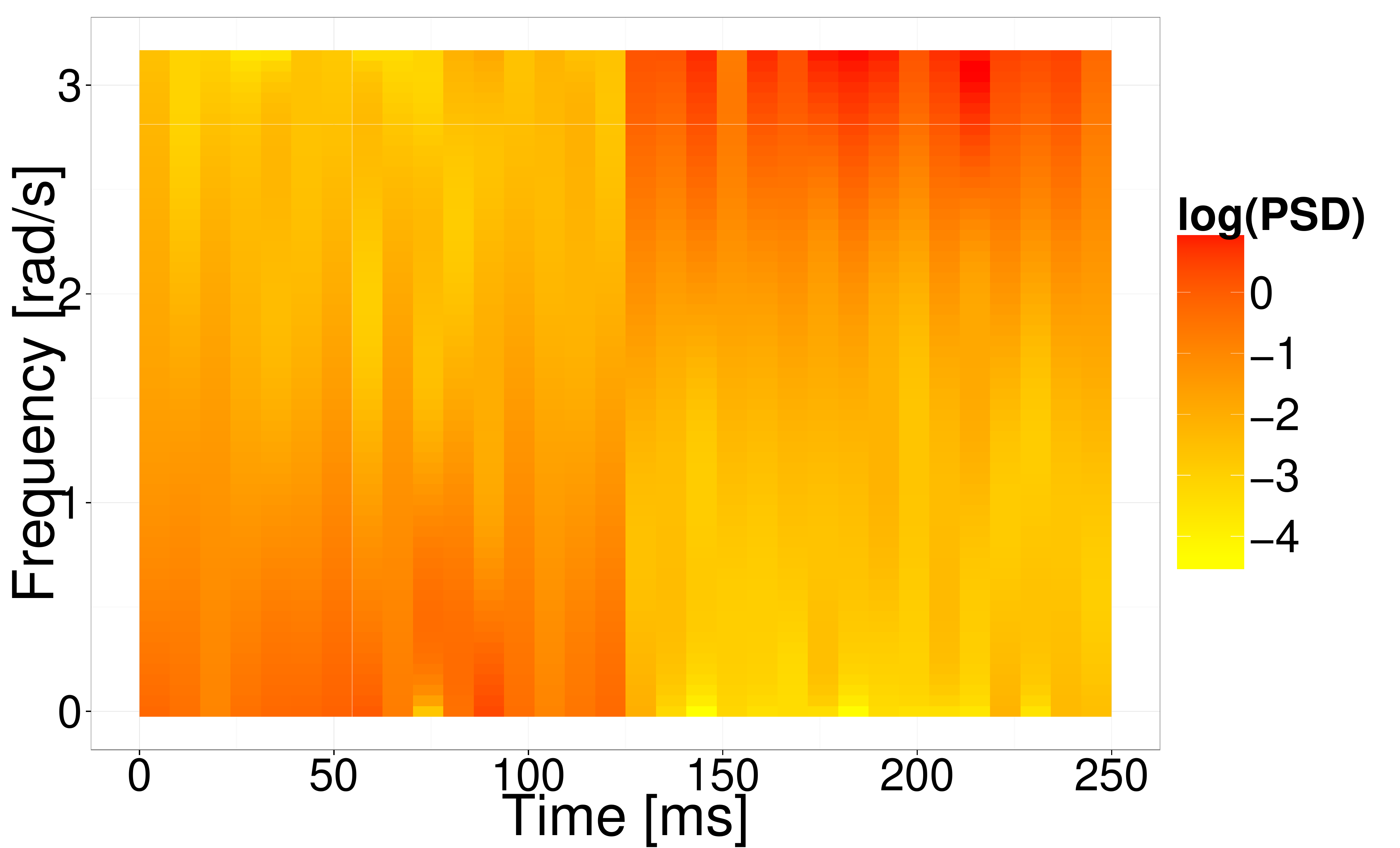}
 \captionof{figure}{Time-frequency map showing the estimated posterior
   median log PSDs for 32 segments of $1 / 4~\mathrm{s}$ of AR(1)
   noise.  The change-point in spectral structure occurs exactly
   halfway through the series.}
\label{fig:spectroAR1_boundary}
\end{center}

It is obvious that a change-point occurs halfway through
Figure~\ref{fig:spectroAR1_boundary}, as there is a sheer change in the
spectral structure at this point between segments 16 and 17.  The
first half of the time-frequency map exhibits stronger low-frequency
behaviour, whereas the second half has more power in the higher
frequencies.

Now consider the case where the change-point occurs during a segment
rather than on the boundary.  Here, let each segment have the same
set-up as before, but instead set $n_1 = 2^{11} - 2^6$ and $n_2 = 2^{11} +
2^6$ such that a change-point occurs halfway through segment 16.  A
time-frequency map of the estimated log PSDs can be seen in
Figure~\ref{fig:spectroAR1_middle}.

\begin{center}
 \includegraphics[width = 1\linewidth]{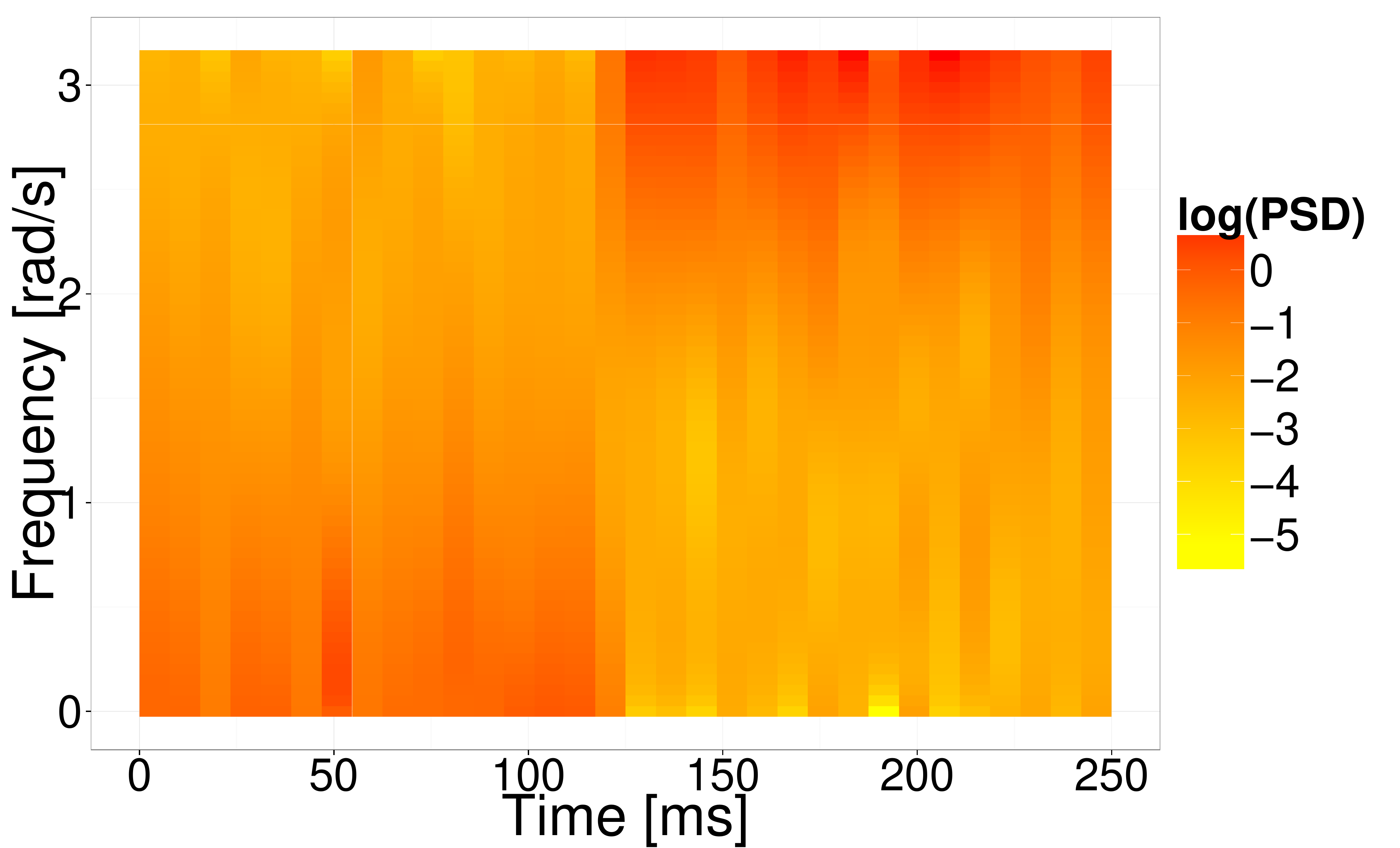}
 \captionof{figure}{Time-frequency map showing the estimated posterior
   median log PSDs for 32 segments of $1 / 4~\mathrm{s}$ of AR(1)
   noise.  The change-point in spectral structure occurs in the middle
   of segment 16 just before the halfway point.}
\label{fig:spectroAR1_middle}
\end{center}

Figure~\ref{fig:spectroAR1_middle} demonstrates that there is a
noticeable change-point roughly halfway through the series.  There is
a smoother transition from one PSD structure to the other than in the
previous example since the true change-point occurs in the middle of a
segment rather than on the boundary.

These examples demonstrate that we can detect potentially unknown
change-points in a time series.  It is important to note that if more
segments are used, the time duration within each segment becomes
smaller, and our accuracy in detecting the change-point increases.
That is, the time at which the change-point occurs becomes more
resolved if the segment durations are smaller.  However, one must also
ensure that the segment durations are long enough for the Whittle
approximation to be valid.

\subsection{Simulated Advanced LIGO Noise}

In this example, we simulate Advanced LIGO noise and embed the
$A1O10.25$ rotating core collapse GW signal from the Abdikamalov
\textit{et al} \cite{abdikamalov:2014} catalogue in it, scaled to an
SNR of $\varrho = 50$.  We assume a one detector set-up, with linearly
polarized GW signal (zero cross polarization).  The Advanced LIGO
sampling rate is $r_s = 2^{14}~\mathrm{Hz}$, with a Nyquist frequency
of $r_* = 2^{13}~\mathrm{Hz}$.  Let $n = 2^{12}$, which corresponds to
quarter of a second of data.

The simulated noise is Gaussian, and coloured by the Advanced LIGO
design sensitivity PSD.  Generating this noise blindly results in a
perfect matching of the end-points and their derivatives, due to the
simplified frequency-domain model.  This is not realistic, since real
data will often not have matching end-points.  In order to make the
noise generation more realistic, we internally generated a longer
frequency-domain series (ten times longer), inverse discrete Fourier
transformed it, and returned a fraction of it with a random starting
point.  This is referred to as ``padding'' the data.

\begin{center}
  \includegraphics[width = 1\linewidth]{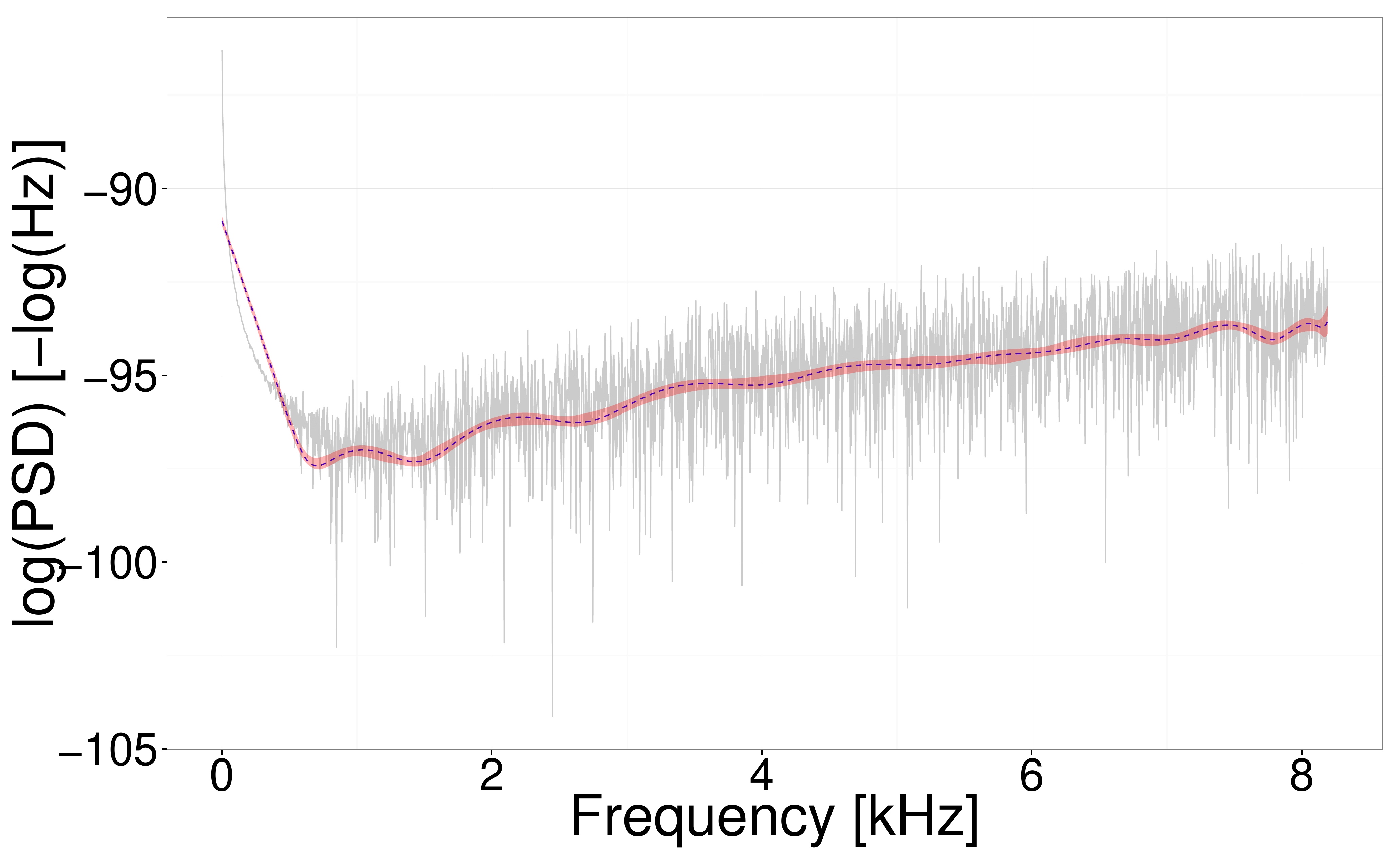}
  \captionof{figure}{Estimated log PSD for simulated Advanced LIGO noise.  90\%
    credible region (shaded pink) and posterior median (dashed blue) overlaid
    with log periodogram (solid grey).}
\label{fig:aLIGO.psd}
\end{center}

Figure~\ref{fig:aLIGO.psd} shows the estimated log PSD and the 90\%
credible region, overlaid with the log periodogram.  The method
performs remarkably well, particularly at higher frequencies.  Even
though we will not be able to resolve frequencies below $\sim
10$--$20~\mathrm{Hz}$ at the Advanced LIGO design sensitivity, it is
still interesting to see how this method performs at lower
frequencies.  Here, the low frequency estimates are slightly off, but
not by much.  We believe this to be due to two factors:
$1/4~\mathrm{s}$ of simulated Advanced LIGO noise is actually a
non-stationary series, and we did not adjust for non-stationarities
(simulated Advanced LIGO data is not stationary for more than
$1/16~\mathrm{s}$ based on the Augmented Dickey-Fuller test,
Phillips-Perron unit root test, and KPSS test); and the Bernstein
polynomial basis functions are notoriously slow to converge to a true
function \cite{powell:1981, shen:2014}.  These factors considered, the
method still provides a reasonable approximation. 

\begin{center}
 \includegraphics[width = 1\linewidth]{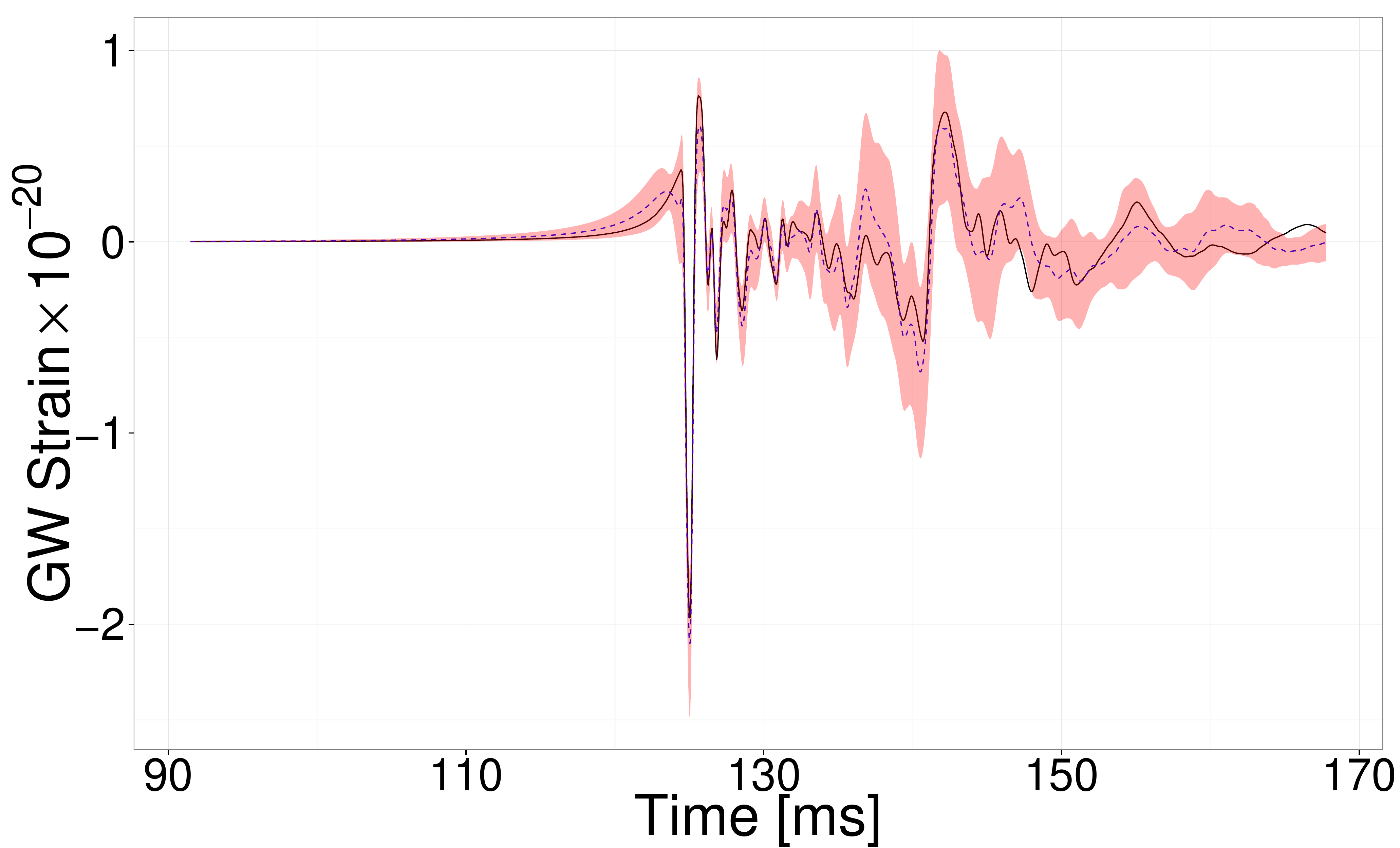}
 \captionof{figure}{Reconstructed rotating core collapse GW signal.  90\%
   credible interval (shaded pink) and posterior mean (dashed blue) overlaid
   with true $A1O10.25$ signal (solid black) from Abdikamalov \textit{et al}
   \cite{abdikamalov:2014} test catalogue.  Signal is scaled to a SNR
   of $\varrho = 50$.}
\label{fig:aLIGO.recon}
\end{center}

The resultant reconstructed GW signal can be seen in
Figure~\ref{fig:aLIGO.recon}.  The estimated signal here is very close
to the true signal during the the collapse and bounce phases, as well
as during the ring-down oscillations.  The 90\% credible region
contains most of the true GW signal.

We chose $d = 25$ PCs to reconstruct a rotating core collapse GW
signal, but this could be too many or too few basis functions.  Model
selection methods similar to \cite{edwards:2014} were not investigated
in the current study, and even though
Figures~\ref{fig:recon.stationary},~\ref{fig:recon.nstat},~and~\ref{fig:aLIGO.recon}
demonstrated good estimates during all phases (including ring-down),
there is a demand for improved reconstruction methods.

We then accommodated for non-stationarities in detector noise by
breaking the series into smaller and locally stationary components,
and looked at the resulting time-varying spectrum.  This can be seen
in Figure~\ref{fig:spectroJ8}.  Rather than choosing $J = 32$ as in
Section~III~E, non-stationarities in the Advanced LIGO noise become
more apparent if we slice the noise series into fewer segments, each
with longer duration.  Instead, consider splitting the data into $J =
8$ equal length segments ($n_j = 2^9$).  Here, the Whittle
approximation is valid, and the segments look locally stationary.

\begin{center}
 \includegraphics[width = 1\linewidth]{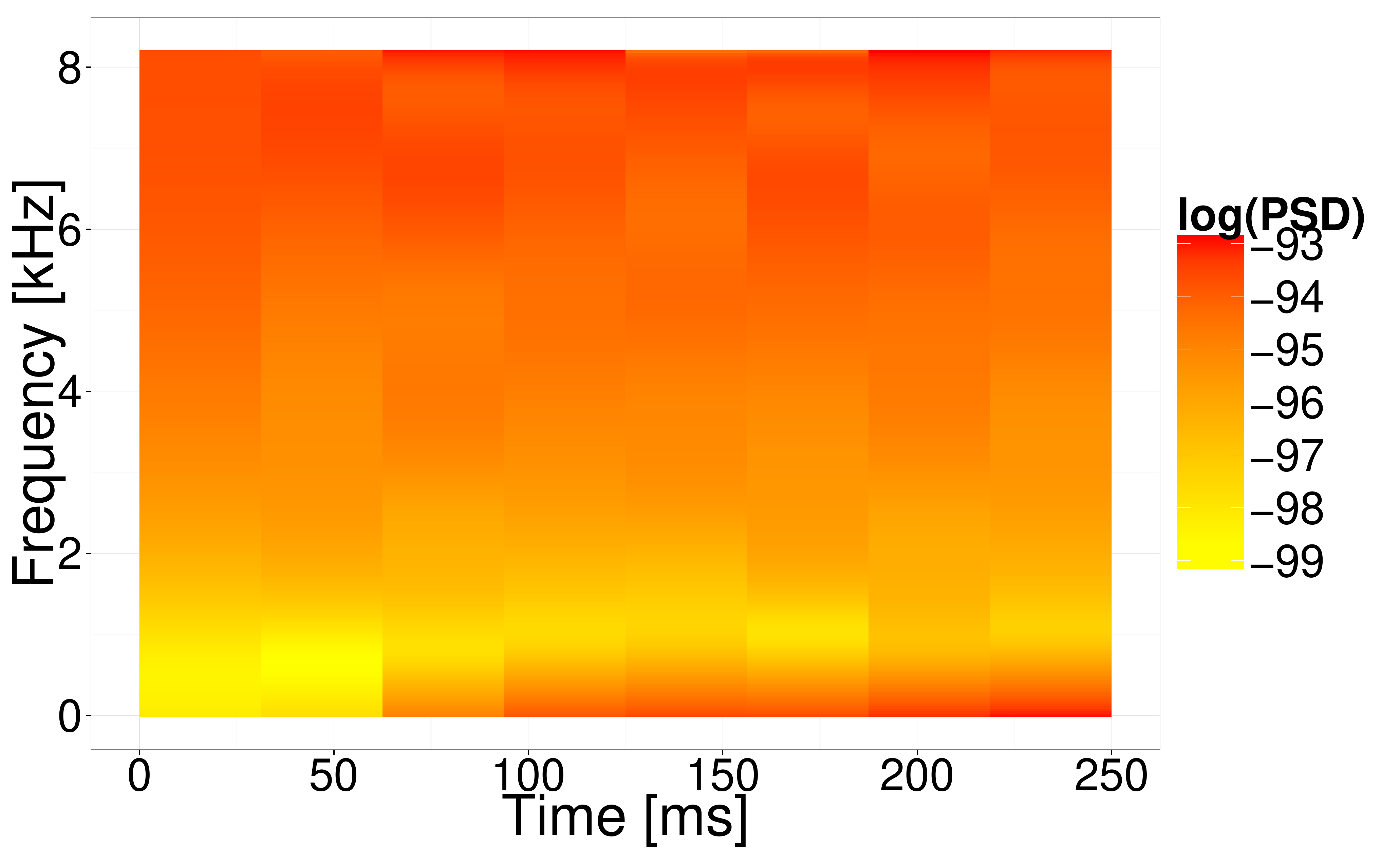}
 \captionof{figure}{Time-frequency map of the estimated time-varying
   noise spectrum for 8 segments of $1 / 4~\mathrm{s}$ simulated
   Advanced LIGO noise.  The posterior median log PSDs for each noise
   segment are used.}
\label{fig:spectroJ8}
\end{center}

Figure~\ref{fig:spectroJ8} illustrates that the Advanced LIGO PSD is
changing over time.  Notice that lower frequencies are gaining more
power over time.  Assuming that each segment is locally stationary
(which should be the case since the duration of each segment is less
than $1/16~\mathrm{s}$), it is important to accommodate for the
changing nature of the PSD since the Choudhuri \textit{et al}
\cite{choudhuri:2004} PSD estimation technique is based on the theory
of stationary processes.  If we did not adjust for non-stationarities,
estimates of astrophysically meaningful parameters could become
biased.

\section{Discussion and Outlook}

This study was motivated by the need for an improved model for PSD
estimation in GW data analysis.  The assumptions of the standard GW
noise model are too restrictive for Advanced LIGO data.  GW data is
subject to high amplitude non-Gaussian transients, meaning that the
Gaussian assumption is not valid.  If the noise model is incorrectly
specified, we could make misleading inferences.  The stationarity
assumption is also not valid, as simulated Advanced LIGO noise is not
stationary for much longer than $1/16~\mathrm{s}$ according to
classical statistical hypothesis tests.  Using off-source data to
estimate the PSD is problematic since the PSD will naturally drift
over time, and is not necessarily the same as on the GW source.

The primary goal of this study was to develop a statistical model that
allows for on-source estimation of the PSD, while making no
assumptions about the underlying noise distribution.  We also wanted a
method capable of accounting for non-stationary noise.  Although we
restricted our attention to GWs from rotating core collapse stellar
events in this paper, our approach is perfectly valid for any GW
signal embedded in noise.

A secondary goal of this paper was to highlight to the GW community the
rich and active area of Bayesian nonparametrics (and semiparametrics).
We believe this framework will be a very powerful toolbox going
forward, particularly in the analysis of GW bursts, since accurate
parametric models for these types of signals are limited.  Further,
our future research efforts regarding rotating core collapse events
involves Bayesian nonparametric regression models to construct GWs
from their initial conditions.  Regularization methods, such as the
Bayesian LASSO \cite{park:2008}, are also being considered.

In this paper, we have assumed linearly polarized GWs to be detected
by one interferometer.  A relatively simple extension of this work is
to include a network of detectors, as well as GWs with non-zero cross
polarization.  Another extension would be to assume an unknown signal
arrival time, as done in \cite{roever:2009, edwards:2014}.  These
extensions can be expected in the second generation of the algorithm.

The noise in our model was assumed to come from all sources, including
detector noise, environmental noise, and statistical noise from
parametric modelling of the signal.  The statistical noise is the
residual difference between the true and fitted signals.  An important
factor to consider was whether statistical noise artificially
dominated the noise.  We do not believe this to be a
dominating contributor to the overall noise.

Since the ``theoretical'' PSD of Advanced LIGO at its design
sensitivity has a very steep decrease at low frequencies until it
reaches a minimum at roughly 230 Hz, it is difficult for our algorithm
to perfectly characterize the shape at low frequencies without
increasing computation significantly.  This is due to the well-known
slow convergence of Bernstein basis functions to a true curve.  That
is, many Bernstein polynomials (on order $k = 1000$) are required to
accurately characterize the PSD of Advanced LIGO.  Compare this to
more well-behaved noise sources, such as those from autoregressive
processes, which require $k < 50$.  We are currently developing a
second generation of this algorithm, using a mixture of B-spline
densities (normalized to the unit interval), rather than Beta
densities.  B-splines have much faster convergence rates than
Bernstein polynomials \cite{powell:1981, shen:2014}.  An additional
benefit of changing the basis functions to B-splines is that, like
BayesLine \cite{bayesline:2014}, we will be able to account for
spectral lines by peak-loading knots at \textit{a priori} known
frequencies that these occur at.  We have left estimation of spectral
lines out of the scope of this paper, but believe that a change of
basis functions from Bernstein polynomials to normalized B-spline
densities could work well.  Another interesting approach would be to
model spectral lines with informative priors using a similar approach
to Macaro \cite{macaro:2010}.

We used non-informative priors in this analysis.  It may be possible
to translate the known shape of the Advanced LIGO design sensitivity
PSD into a prior.  This may also aid in improving PSD estimates at
lower frequencies.

We discussed a simplified method for estimating the time-varying PSD
of non-stationary noise.  Our approach assumed that a time series is
split into equal length segments, and at known times.  We demonstrated
that it is possible to identify change-points in a time series and its
spectrum using this method, and that there is no need to estimate the
locations of the segment splits.  Thus, a fixed grid of known segment
placements suffices, and no RJMCMC is required.  RJMCMC would have
slowed the algorithm down significantly, and created an entire new set
of complications.

There is much work to be done on PSD estimation.  As the Advanced LIGO
and Advanced Virgo interferometers swiftly approach design
sensitivity, it is important that we continue to focus not only on
parameter estimation techniques, but also on modelling detector noise.
PSD estimation is as important as parameter estimation, since we want
to make honest statements about our observations based on rigorous
statistical theory.  It is hoped that in the near future, we can
converge on a PSD estimation method that is less strict than the
standard noise model, works well on real detector data, and is based
on good statistical theory.  We believe that the methods presented in
this paper are definitely a step in the right direction.

\appendix

\section{Bernstein Polynomials and the Beta Density}

To define the Bernstein polynomial, we first need to discuss the
Bernstein \textit{basis} polynomials.  There are $k + 1$ Bernstein
basis polynomials of degree $k$, having the following form
\begin{equation}
b_{j, k}(x) = {k \choose j} x^j (1-x)^{k-j}, \quad j = 0, 1, \ldots, k.
\end{equation}

A Bernstein polynomial is the following linear combination of
Bernstein basis polynomials
\begin{equation}
B_k(x) = \sum_{j = 0}^k \beta_j b_{j, k}(x),
\end{equation}
where $\beta_j$ are called the Bernstein coefficients.

As mentioned in Section~II~C, the Bernstein polynomial prior is a
finite mixture of Beta probability densities.  We use the following
parameterization for the Beta probability density function
\begin{eqnarray}
f(x | \alpha, \beta) &=& \frac{\Gamma(\alpha + \beta)}{\Gamma(\alpha)\Gamma(\beta)}x^{\alpha - 1}(1-x)^{\beta-1}, \\
&\propto& x^{\alpha-1}(1-x)^{\beta-1},
\end{eqnarray}
where $x \in (0, 1)$, the shape parameters are positive real numbers
(i.e., $\alpha > 0$ and $\beta > 0$), and $\Gamma(.)$ is the gamma
function defined as the following improper integral
\begin{equation}
  \label{eq:gamma}
  \Gamma(u) = \int_0^{\infty} x^{u-1}\exp(-x)\mathrm{d}x.
\end{equation}

\section{The Dirichlet Distribution, Dirichlet Process, and Stick-breaking Construction}

The Dirichlet distribution is a multivariate generalization of the
Beta distribution (defined in Appendix~A) with a probability density
function defined on the $K$-dimensional simplex
\begin{equation}
  \label{eq:simplex}
  \Delta_K = \left\{(x_1, \ldots, x_K): \quad x_i > 0, \quad \sum_{i = 1}^K x_i = 1 \right\}.
\end{equation}

The probability density function of the Dirichlet distribution is defined as
\begin{equation}
  \label{eq:dirichlet}
  f(\mathbf{x} | \boldsymbol\alpha) = \frac{\Gamma\left(\sum_{i = 1}^K \alpha_i \right)}{\prod_{i = 1}^K \Gamma(\alpha_i)} \prod_{i = 1}^K x_i^{\alpha_i -1}, 
\end{equation}
where $\alpha_i > 0, i = 1, \ldots, K$.

The Dirichlet process is an infinite-dimensional generalization of the
Dirichlet distribution.  It is a probability distribution on the space
of probability distributions, and is often used in Bayesian inference
as a prior for infinite mixture models.  One of the many
representations of the Dirichlet process is Sethuraman's
stick-breaking construction \cite{sethuraman:1994, BNP:2010}.  This is
useful for implementing MCMC sampling algorithms.
 
Let $G \sim \mathrm{DP}(M, G_0)$, where $G_0$ is the center measure,
and $M$ is the precision parameter (larger $M$ implies a more
precise prior).  The Sethuraman representation is
\begin{eqnarray}
  \label{eq:stickbreaking}
  G &=& \sum_{i = 1}^{\infty} p_i\delta_{Z_i}, \\
  p_i &=& \left(\prod_{j = 1}^{i - 1} \left(1 - V_j\right)\right)V_i, \\
  Z_i &\sim& G_0, \\
  V_i &\sim& \mathrm{Beta}(1, M).
\end{eqnarray}

Consider a stick of unit length.  The weights $p_i$ associated with
points $Z_i$ can be thought of as breaking this stick randomly into
infinite segments.  Break the stick at location $V_1 \sim
\mathrm{Beta}(1, M)$, assigning the mass $V_1$ to the random point
$Z_1 \sim G_0$.  Break the remaining length of the stick $1 - V_1$ by
the proportion $V_2 \sim \mathrm{Beta}(1, M)$, assigning the mass $(1
- V_1)V_2$ to the random point $Z_2 \sim G_0$.  At the
$i^{\mathrm{th}}$ step, break the remaining length of the stick
$\prod_{j = 1}^{i - 1}(1 - V_j)$ by the proportion $V_i \sim
\mathrm{Beta}(1, M)$, assigning the mass $\left(\prod_{j = 1}^{i -
    1}\left(1 - V_j\right)\right) V_i$ to the random point $Z_i \sim
G_0$.  This process is repeated infinitely many times.

\section{Demonstration of Posterior Consistency}

It was proved in \cite{choudhuri:2004} that under very general
conditions on the prior, the PSD estimation method used in this paper
has the property of posterior consistency.  We provide an illustrative
example of this in Figure~\ref{fig:postconsist}.\\~\\

\begin{center}
\includegraphics[width=1\linewidth]{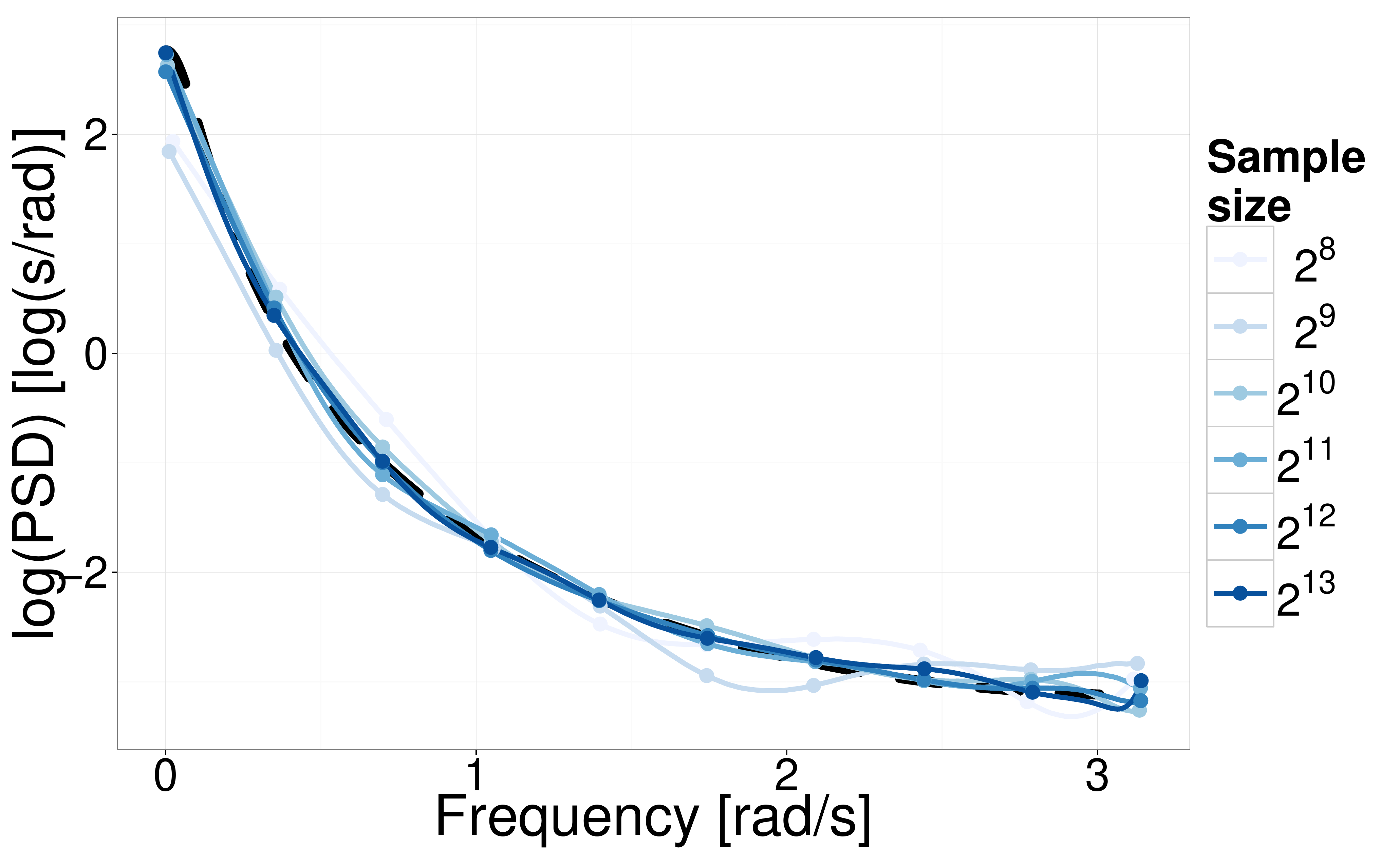}
\captionof{figure}{Illustration of posterior consistency.  The true
  log PSD (dashed black) overlaid with point-wise posterior median log
  PSDs of varying sample sizes.}
\label{fig:postconsist}
\end{center}

We generated AR(1) processes (with $\rho = 0.9$ and Gaussian white
noise) of varying sample sizes, and compared their performance.  It
can be seen in Figure~\ref{fig:postconsist} that as the sample size of
the time series increases, the point-wise posterior median log PSD gets
closer to the true log PSD, thus demonstrating posterior consistency.

\begin{acknowledgements}
  We thank Neil Cornish for a thorough reading of the manuscript and
  for providing insightful comments, Claudia Kirch and Sally Wood for
  helpful discussions on stationary and locally stationary time
  series, and Blake Seers for advice on data visualization.  We also
  thank the New Zealand eScience Infrastructure (NeSI) for their high
  performance computing facilities, and the Centre for eResearch at
  the University of Auckland for their technical support. RM's and
  ME's work is supported by the University of Auckland staff research
  grant 420048358, and NC's work by NSF grants PHY-1204371 and
  PHY-1505373.  This paper has been given LIGO Document Number
  P1500057.  All analysis was conducted in \textsf{R}, an open-source
  statistical software available on \textsf{CRAN}
  (cran.r-project.org).  We acknowledge the \textsf{ggplot2},
  \textsf{grid}, \textsf{coda}, and \textsf{bspec} packages.
\end{acknowledgements}


\bibliographystyle{apsrev4-1}


\end{document}